\documentclass[iop]{emulateapj}

\begin{document}
  
  \title{A Stochastic Model for the Luminosity Fluctuations of Accreting Black Holes}

  \author{Brandon C. Kelly\altaffilmark{1,2,3}, Ma{\l}gorzata Sobolewska\altaffilmark{3}, Aneta
    Siemiginowska\altaffilmark{3}}

  \altaffiltext{1}{bckelly@cfa.harvard.edu}
  \altaffiltext{2}{Hubble Fellow}
  \altaffiltext{3}{Harvard-Smithsonian Center for Astrophysics, 
      60 Garden St, Cambridge, MA 02138}
  
  \begin{abstract}
In this work we have developed a new stochastic model for the
fluctuations in lightcurves of accreting black holes. The model is
based on a linear combination of stochastic processes and is also 
the solution to the linear diffusion equation perturbed by a
spatially correlated noise field. This allows flexible modeling of the
power spectral density (PSD), and we derive the likelihood function
for the process, enabling one to estimate the parameters of the
process, including break frequencies in the PSD.  Our statistical
technique is computationally efficient, unbiased by aliasing and red
noise leak, and fully accounts for irregular sampling and measurement
errors.  We show that our stochastic model provides a good
approximation to the X-ray lightcurves of galactic black holes, and
the optical and X-ray lightcurves of AGN. We use the estimated time
scales of our stochastic model to recover the correlation between
characteristic time scale of the high frequency X-ray fluctuations and
black hole mass for AGN, including two new `detections' of the time
scale for Fairall 9 and NGC 5548. We find a tight anti-correlation
between the black hole mass and the amplitude of the driving noise
field, which is proportional to the amplitude of the high frequency
X-ray PSD, and we estimate that this parameter gives black hole mass
estimates to within $\sim 0.2$ dex precision, potentially the most
accurate method for AGN yet. We also find evidence that $\approx 13\%$
of AGN optical PSDs fall off flatter than $1 / f^2$, and, similar to
previous work, find that the optical fluctuations are more suppressed
on short time scales compared to the X-rays, but are larger on long
time scales, suggesting the optical fluctuations are not solely due to
reprocessing of X-rays.
  \end{abstract}
  
  
  \section{INTRODUCTION}

  \label{s-intro}

  Both galactic black holes (GBH) and active galactic nuclei (AGN),
being black hole accretion systems, share a number of similarities in
their spectroscopic and variability features despite being separated
by several decades in mass. The major ingredients of the two systems
appear to be an optically thick, geometrically thin accretion disk
emitting thermal radiation, and an optically thin hot `corona'
emitting X-ray emission with a power-law spectrum. Because the
temperature of the disk scales with the black hole mass as $kT \propto
M_{BH}^{-1/4}$, the thermal disk emits in the soft X-rays for GBHs,
and in the optical/UV region for AGN. The source and geometry of the
corona emission is not known, but likely involves inverse compton
emission from a hot electron plasma \citep[e.g.,][]{shap76,haardt91}
with a possible synchrotron contribution from a jet
\citep{markoff01,markoff05}. Both GBHs and AGN follow a correlation
between the black hole mass, radio luminosity, and X-ray luminosity
\citep{merloni03,kord06a,yuan09,gult09b}, leading some to suggest that
the systems are largely self-similar, at least in low accretion rate
states, differing only in mass and environment
\citep[e.g.,][]{falcke04}.

  However, the comparison between GBHs and AGNs is not
straightforward, as GBHs are known to exist in different `states',
with the same source seen cycling between the various states
\citep[for a review see][]{rem06,belloni10}. The spectral and timing
properties of GBHs are different in the different states. The most
commonly observed states are the hard, soft, and very high state
(VHS). In the hard state, the spectrum is dominated by the power-law
hard X-ray emission, while in the soft state the X-ray flux increases
and the spectrum contains a significant thermal disk component. In the
VHS, the X-ray flux is very high and the spectrum is intermediate in
hardness between the soft and hard states. The power spectral density
(PSD, $P(f)$) of the X-ray fluctuations can be reasonably approximated
by bending power-laws, $P(f) \propto 1 / f^\alpha$. For the soft and
intermediate states, the power-law component contributes to the
majority of the variability \citep[e.g.,][]{done05,sob06}, while for
the hard state the disk blackbody component dominates the soft X-ray variability,
at least on time scales $> 1$ sec \citep{wilk09}. The PSD at high
frequencies has a logarithmic slope $\alpha \gtrsim 2$, flattening to
a slope of $\alpha \sim 1$ below some break frequency. For GBH in the
hard state and VHS, there is an additional flattening to $\alpha = 0$
below a second lower break frequency, with the break frequencies being
higher in the VHS. The accretion rate increases as one moves from the
hard state to the VHS.

  For AGN the situation is less clear, due to the fact that the time
scales involved for the state transitions are expected to scale
upwards with black hole mass. As such, AGN, with their supermassive
black holes (SMBH), have not been observed to unabmiguously undergo
state transitions. Many AGN display SEDs more characteristic of soft
state GBH, in that there is a strong thermal component from the disk
in the optical/UV region. \citet{sob09a} suggest, based on comparisons
of the AGN and GBH SEDs, that most AGN in current surveys are in the
VHS \citep[see also][]{sob11}. Almost all of the $\sim 10$ AGN with high quality X-ray
lightcurves exhibit PSDs similar to those seen in Cyg X-1 in the soft
state \citep[e.g.,][]{mark03,mchardy06}. The only exception is Akn
564, which exhibits a second low-frequency break
\citep{arev06a,mchardy07}. The high accretion rate \citep[$\dot{m}
\sim 1$,][]{rom04} led \citet{mchardy07} to suggest that this source
is analagous to the VHS. However, the average 2--10 keV photon index
of these sources is $\Gamma < 2$ \citep{pap09,sob09b}, which resembles
the value observed for GBHs in the hard state; thus, there is some
descrepancy between the spectral and timing classifications. The break
frequencies for both GBHs and AGN are observed to anti-correlate with
black hole mass \citep{uttley05a,mchardy06}, further strengthening the
similarity between GBHs and AGN. The anti-correlation between the
break frequency and black hole mass has typically been interpreted as
being driven by a correlation between the size of the X-ray emitting
region and black hole mass. A number of studies have also found an
anti-correlation between the amplitude of X-ray variability and
$M_{BH}$
\citep[e.g.][]{lu01,bian03,pap04,nik04,oneill05,gier08,zhou10}. The
absence of AGN analogous to the hard state suggests that selection
effects may be at work, as all GBHs with accretion rates less than a
few per cent of Eddington are hard state objects
\citep[e.g.,][]{maccarone03}, while most AGN surveys have only been
able detect objects radiating at $L / L_{Edd} \gtrsim 0.1$ in large
numbers \citep[e.g.,][]{vest04,trump09a,kelly10}.

  While studies of AGN variability have many complications, they do
have the advantage that the disk and corona emission can be cleanly
separated. In contrast, for GBH in the soft state the seperation is
more difficult as both components emit in X-rays, and the disk
component does not exhibit significant variability
\citep[e.g.,][]{churaz01}. Therefore, studies of the disk emission
variability are more easily carried out for AGN. Previous studies of
AGN optical variability using well-sampled lightcurves have found that
the optical variations have dispersions of $\sim 10$--$20\%$
\citep{kbs09,macleod10a}, with the X-rays varying more on the shorter
time scales \citep[e.g.,][]{ulrich97,czerny03,smith07,arev09}, and
sometimes also on the longer time scales \citep{breedt10}. In addition,
the optical PSD can be well described by a power law form $P(f) \propto 1
/ f^2$ \citep{giveon99,collier01,czerny03}, or, when the lightcurve is
long enough, by a Lorentzian centered at zero
\citep{czerny99,kbs09,koz10,macleod10a}; i.e., the PSD is $P(f) \propto
1 / f^2$, flattening to a constant below some break frequency. Similar
to the X-ray PSD, the break frequency of the optical PSD increases
with black hole mass \citep{collier01,kbs09,macleod10a}, although not
as steeply.

  While variability studies of AGN and GBHs have been important for
understanding the similarities between these two classes of objects,
variability studies are important for other reasons as well. For one,
variability is one of the only observational tools available for
probing the disk viscosity, as the time scales of variability are
expected to depend on viscosity, while the SED is not
\citep{siem89,star04,frank02}. Understanding the disk viscosity is
important for understanding the transfer and removal of angular
momentum in the disk, which is fundamental to an understanding of the
accretion flow. Variability is also a potentially important
observational tool for constraining the geometry of the corona, with,
for example, studies of quasi-periodic oscillations being consistent
with the disk evaporating into a hot inner flow at lower accretion
rates \citep[e.g.,][]{cui99,rossi04,done07}, and the origin of the
variability in the QPOs being in the hot corona in the soft state and
the disk in the hard state \citep[e.g.,][]{sob06}. This is also
consistent with the finding of \citet{wilk09} who showed that the disk
is responsible for much of the variability in the hard state; however,
there is still much work to be done, as it has not been conclusively
shown that the source of variability originates in the same region as
the emission. In addition, variability may be the most effective
observational discriminator between the different accretion states,
and thus may provide evidence of radiatively inefficient accretion
flows (RIAFs) in AGN, which are believed to be associated with the
hard state, should a hard state exist for AGN. A number of RIAF
candidates exist \citep[e.g.,][]{ho99,quat99,trump09b}, but their
spectral/timing state is unknown. Studies and confirmation of AGNs in
the hard state are particularly important, as the hard state is
associated with jet production in GBHs
\citep[e.g.,][]{fender01,fender04,kord06b}, and mechanical feedback
from these jets plays an important role in heating intracluster gas
and regulating the growth of massive galaxies
\citep{croton06,bower06,sijacki07}.

Time series exhibiting PSDs of a $1 / f$ type are known as
'long-memory' processes, and there is an extensive literature on them
\citep[e.g., a good reference is][]{palma07}. Almost all previous
studies of variability in GBHs and AGN have relied on non-parameteric
techniques, such as the periodogram or the structure
function. However, there are a number of known difficulties in
estimating the PSD or structure function non-parameterically
\citep[e.g., see][]{vaughan03,pessah07}. For one, the empirical
estimate of the PSD, known as the periodogram, suffers from windowing
effects due to the finite sampling of the lightcurve. These windowing
effects include red noise leak and aliasing, which are caused by power
leaking into the periodogram from time scales longer and shorter than
the maximum and minimum time scales probed by the lightcurve,
respectively. Red noise leak and aliasing distort the periodogram,
making it potentially difficult to relate the observed periodogram to
the true intrinsic PSD. Moreover, irregular sampling further distorts
the periodogram, although this distortion can be alleviated through
the use of the Lomb-Scargle periodogram
\citep{lomb76,scargle82,horne86,zech09,vio10}. The structure function
is not immune to these problems and can similarly be distorted, making
its interpretation difficult \citep{emman10}. These distortion
problems can be significant for AGN especially, as their observed flux
is often much fainter than GBHs and their lightcurves tend to be more
irregular and sparsely sampled.

  Motivated by these problems, \citet[][see also
\citet{done92}]{uttley02} developed a powerful Monte Carlo technique
for estimating the underlying PSD which accounts for the distorting
effects of finite, and possibly irregular, sampling. The basic idea
behind the technique is to use Monte Carlo simulations to calculate
the expected value of the periodogram as a function of the true
underlying PSD, and then to find the true PSD which minimizes a
$\chi^2$ goodness of fit measure between the observed periodogram and
the expected one. Construction of confidence regions may be obtained
through the procedure outlined by \citet{mueller09}.  This technique
has the advantage that it may be used to fit any arbitrary
PSD. However, it also has two disadvantages. First, it is
computationally intensive, especially when constructing confidence
regions. Second, the $\chi^2$ goodness of fit statistic used to fit
the PSDs ignores the covariance in the periodogram among the frequency
bins, and is not proportional to the log-likelihood of the
periodogram. As such, minimization of the $\chi^2$ statistic of
\citet{uttley02}, while effective, is unlikely to be the most
efficient means of constraining the underlying PSD. An alternative
Bayesian approach based on an approximation to the likelihood function
of the periodogram has been developed by \citet{vau10}. An additional
likelihood-based approach has been developed by \citet{miller10} with
the difference being that the likelihood function is calculated in the
time domain. Unfortunately these two alternative approaches do not
completely incorporate the distortion in the PSD due to finite
sampling of the lightcurve, such as that caused by red noise leak,
although in principle the PSD model used by \citet{miller10} can be
modified to correct for this. Indeed, it is very difficult to derive
an analytic expression for the likelihood function of the periodogram
for a finite and irregularly sampled lightcurve, and in many cases may
be impossible.

  An alternative and complementary approach to Fourier-based techniques is
to model the lightcurve as a parameterized stochastic process, with
the parameters of the model being related to the underlying PSD (or
rather, the PSD being a function of the parameters of the
process). Under this approach, the parameters of the model are
estimated directly from the lightcurve itself; no Fourier transforms
are performed, and thus there is no spectral distortion. Moreover,
modeling the lightcurve in the time domain potentially can provide
further insight on features in the power spectrum, such as break
frequencies, as it is not always apparent how to interpret the
PSD. Time-domain modeling was advocated for by \citet[][hereafter
KBS09, see also \citet{scargle80} and \citet{ryb92}]{kbs09} within the
context of estimating the characteristic time 
scale of AGN optical variability (or equivalently, the break
frequency of the optical PSD), and they developed a Bayesian approach for estimating the
parameters. KBS09 modeled AGN optical lightcurves as 
Gaussian Ornstein-Uhlenbeck (OU) process, the power spectrum of which is a
Lorentzian centered at zero, and showed that this process is
consistent with the optical lightcurves of AGN for their
sample. Subsequent work has confirmed that the OU process provides a
good description of AGN optical variability, at least on time scales
between a few days and several years \citep{koz10,macleod10a}, and of
blazar sub-mm variability \citep{strom10}. Moreover, the OU process provides a framework in which to measure the time lags between the AGN broad line region and optical continuum \citep{zu10} and select quasars \citep{koz10,butler10,macleod10b}.

  In this work we extend the method of KBS09 to enable more flexible
modeling of the PSD of accreting black hole systems. We model the
lightcurves as a linear expansion of OU processes, which enable us to
approximate many of the features seen in the PSDs of GBHs and AGN,
particularly the bending power-law forms. We derive the likelihood
function for this statistical model and perform statistical inference
within a Bayesian framework, thus enabling one to calculate the
probability distribution of the parameters, such as the break
frequencies, given the data. Because our method is based on the
likelihood function, it uses all of the information in the
data. Furthermore, it fully accounts for measurement errors,
irregularly sampling, red noise leak, and aliasing. Fitting is
performed on the entire lightcurve simultaneously regardless of the
sampling, thus enabling one to easily combine lightcurves obtained
from different instruments and sampling time scales. These properties
make our technique particularly attractive for obtaining constraints
on PSD features for poorly sampled lightcurves of faint objects, as
our Bayesian method uses all of the information in the
data. Calculation of the likelihood function is computationally
efficient, and we are able to calculate a maximum-likelihood estimate
of the PSD in under a minute\footnote[1]{The calculation was done on a
lightcurve with $\sim 3000$ data points in IDL on a Mac Pro with two
3.2 GHz Quad-Core Intel Xeon processors}. The primary disadvantage of
our method is that, while flexible, it cannot fit arbitrary PSDs; for
this, one should use the method of \citet{uttley02}, or a combination
of the two techniques. 

  While the OU process, and similarly the mixture of OU processes, is
useful for fitting PSDs, and thus quantifying variability, it is not
always clear how to interpret the best-fit OU process astrophysically,
or the PSD in general. \citet{tit07} studied the linear diffusion equation for an
accretion disk with a driving noise term as a model 
for the PSDs of accreting black holes, showing that features in the
PSD depend on the viscosity of the accretion flow. In this work we will show how the mixture
of OU processes arises as the solution to the linear diffusion
equation, perturbed by a random spatially-correlated white noise
field. This interpretation
of the mixed OU process may be considered to be among the
`perturbation' class of astrophysical models for variability of GBHs
and AGN, with the variability arising from small random accretion rate
perturbations that propagate inwards through the accretion flow,
making the observed variations the product of perturbations that
occurr at larger radii
\citep[e.g.,][]{lyub97,king04,arev06b,janiuk07,tit07}. The
perturbation class of models for variability also explains the
correlation between flux and absolute RMS variability seen in both GBHs and AGN
\citep[e.g.][]{uttley01}, which \citet{uttley05b} show is very well
approximated as being due to a Gaussian process on the logarithmic
scale. 

Finally, we note that the perturbation class of models for variability
is only applicable to the broad-band flickering noise seen in GBH and
AGN lightcurves, and we do not address the more catastrophic, and
seemingly not stochastic, changes that have been observed in the
lightcurves of GRS 1915+105 \citep[e.g.,][]{belloni00}. These may
occur due to accretion disk instabilities, such as radiation pressure
instabilities \citep[e.g.,][]{shak76,lightman74,czerny09}, or the
thermal-viscous ionization instability
\citep{lin86,siem96,janiuk04}. Furthermore, we stress that the
astrophysical interpretation of the mixed OU process as a model for
the fluctuations from GBHs and AGN is based on the linear diffusion
equation, which is surely an oversimplification, and MHD simulations
must be performed for studying more physically motivated models for
variability of accreting black holes, as done by, e.g.,
\citet{armitage03}, \citet{schnitt06}, \citet{mosc07}, \citet{reynolds09}, and
\citet{noble09}. Rather, we study the mixed OU process as a solution
to the stochastic linear diffusion equation in order to provide a
guide for interpreting the best-fit model and PSD features of
lightcurves, both real and simulated, as analytical solutions may be
obtained.

  The format of the paper is as follows: In \S~\ref{s-smodel} we
  describe the mixed OU process statistical model for the lightcurves
  of GBHs and AGN. In \S~\ref{s-oudiff} we show that the mixed OU
  process is generically the solution to the linear diffusion
  equation. In \S~\ref{s-likhood} we derive the likelihood function of
  the mixed OU process parameters, and posterior probability
  distribution of the parameters given an observed lightcurve. In
  \S~\ref{s-apply} we apply the mixed OU process to an X-ray
  lightcurve of Cygnus X-1 in the low/hard state, the X-ray
  lightcurves of 10 local well-studied AGN, and the optical
  lightcurves of a sample of AGN. Finally, in \S~\ref{s-summary} we
  summarize our results.

  \section{THE STATISTICAL MODEL}

  \label{s-smodel} 

  \subsection{The Ornstein-Uhlenbeck Processes}

  \label{s-ou}

  In this section we give a brief overview of the properties of the OU
  process that are relevant for our work. Further
  details can be found in \citet[][see also \citet{gill96} and \citet{gard04}]{kbs09},
  who refer to this process as a first-order continuous autoregressive
  process. The OU process, $X(t)$, is a simple stochastic process by
  which the quantity of interest (say, the logarithm of the
  luminosity or accretion rate) responds to 
  an input noise process with an exponential decay to its
  mean. Mathematically, the OU process is defined by the following
  stochastic differential equation:
  \begin{equation}
    dX(t) = -\omega_0 (X(t) - \mu) dt + \varsigma dW(t), \ \ \omega_0,
    \varsigma > 0.
    \label{eq-ou}
  \end{equation}
  The parameters of the process are the characteristic frequency,
  $\omega_0$, the mean value of the process, $\mu$, and the amplitude
  of the driving noise process, $\varsigma$. The term $\varsigma^2$ has units of ${\rm RMS^2\ sec^{-1}}$, and thus $\varsigma$ gives the rate at which variability power is injected into the stochastic process $X(t)$. The term $W(t)$ denotes a Wiener
  process (i.e., a Brownian motion), and its derivative is white
  noise. Although in this work we will focus on the special case when
  $W(t)$ is a Wiener process, the results outlined here for the OU
  process, such as the form of its PSD, are valid for a more general
  class of stochastic processes called L\'{e}vy processes. For
  completeness, we give a brief description of L\'{e}vy processes in
  the appendix. It is apparent from
  setting $\varsigma = 0$ in Equation 
  (\ref{eq-ou}) that $X(t)$ decays to its mean value with an
  $e$-folding time scale of $\tau = 1 / \omega_0$. The time scale
  $\tau$ is often called the relaxation time scale of the process; in
  this work we will refer to $\tau$ as the characteristic time scale
  of the process, as $\tau$ is related to the break frequency in the
  power spectrum of $X(t)$.

  When the driving noise, $dW(t)$, is white, the OU process is stationary and
  Markov. A stationary process is one whose joint probability
  distribution does not change when shifted in time, and a Markov
  process is one whose future states only depends on the current
  state. In addition, if $W(t)$ has zero mean and unit variance the 
  autocovariance function of the OU process is 
  \begin{equation}
    R_{OU}(t) = \frac{\varsigma^2 \tau}{2} e^{-|t| / \tau}.
    \label{eq-Rou}
  \end{equation}
  It follows from setting $t = 0$ in Equation (\ref{eq-Rou}) that the variance of the OU
  process is $\varsigma^2 \tau / 2$, and that the autocorrelation
  function has an exponential decay with $\tau = 1 / \omega_0$ being the
  decorrelation time scale. The power spectrum of the OU process is
  given by the Fourier transform of the autocovariance function:
  \begin{eqnarray}
    P_{OU}(\omega) & = & \frac{1}{2\pi} \int_{-\infty}^{\infty}
    e^{-i\omega t} R_{OU}(t)\ dt \nonumber \\
    & = & \frac{\varsigma^2}{2\pi} \frac{1}{\omega_0^2 + \omega^2}.
    \label{eq-Pou}
  \end{eqnarray}
  The power spectrum of an OU process is flat for frequencies $\omega
  \ll \omega_0$, and decays as $1 / \omega^2$ for frequencies $\omega
  \gg \omega_0$. Hence the association of $\tau$ as a characteristic
  time scale of the process: the OU process, $X(t)$, resembles white
  noise on time scales long compared to $\tau$, and resembles red noise
  on time scales short compared to $\tau$. Note that if one calculates
  the power spectrum in terms of ordinary frequencies instead of angular
  frequencies, then $\tau = 1 / (2 \pi f_0)$, where $f_0 = \omega_0 /
  (2\pi)$. This implies that simply fitting the break in the power
  spectrum as a function of ordinary frequency $f$ will overestimate
  $\tau$ by a factor of $2 \pi$.

  \subsection{Mixtures of Ornstein-Uhlenbeck Processes}

  \label{s-supou}

  We can build on the OU process to develop more flexible models that
  enable modeling of more complex power spectra. In this work we
  consider a mixture of OU processes with independent driving noises
  for constructing a process with a power spectrum of the form
  exhibited by the X-ray lightcurves of GBH and AGN. We define a
  discrete mixture of OU processes as 
  \begin{equation}
    Y_M(t) = \mu + \sum_{i=1}^M c_i X_i(t).
    \label{eq-mixou_disc}
  \end{equation}
  Here, $c_1, \ldots, c_M$ is a set of mixing weights, and $X_1(t),
  \ldots, X_M(t)$ is a set of OU processes with characteristic
  frequencies $\omega_1,\ldots,\omega_M$ and driving noise amplitudes
  $\varsigma_1,\ldots,\varsigma_M$. A continuous version of Equation
  (\ref{eq-mixou_disc}) is described in the Appendix, although we do
  not use it in this work. The 
  autocovariance function of $Y_M(t)$ is
  \begin{equation}
    R_{Y,M}(t) = \sum_{i=1}^M \frac{c_i^2 \varsigma_i^2 \tau_i}{2}
    e^{-|t| / \tau_i},
    \label{eq-Rmixou_disc}
  \end{equation}
  where we have used the fact that $\tau_i = 1 / \omega_i$. The sum of
  exponentials in Equation (\ref{eq-Rmixou_disc}) falls off more
  slowly than a single exponential 
  function, and therefore the autocorrelation function of the mixed OU
  process falls off more slowly than the autocorrelation function of
  the individual OU processes. In other words, the mixed OU process
  exhibits longer range dependency than a single OU process, and
  therefore is better suited for modeling lightcurves with long time
  scale dependencies. 

  The power spectrum of $Y_M(t)$ is
  \begin{equation}
    P_{Y,M}(t) = \sum_{i=1}^M \frac{c_i^2 \varsigma_i^2}{2\pi} \frac{1}{\omega_{i}^2 + \omega^2}.
    \label{eq-Pmixou_disc}
  \end{equation}
  From Equations (\ref{eq-Rmixou_disc}) and (\ref{eq-Pmixou_disc}) it
  is apparent that the contribution of an individual OU process to the variability amplitude of a lightcurve depends on the product $c_i
  \varsigma_i$, and thus the two parameters are degenerate. Therefore,
  for simplicity in this work we only consider the case where all of
  the individual OU processes have the same value of $\varsigma$. 

\begin{figure}
  \begin{center}
    \scalebox{1.0}{\rotatebox{0}{\plotone{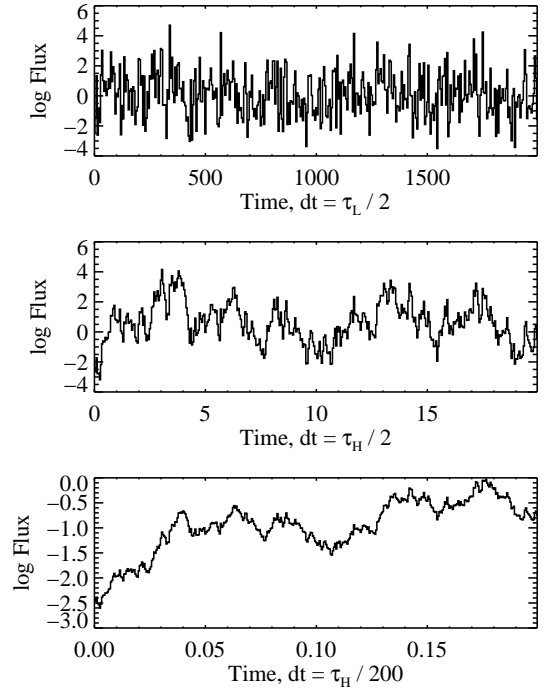}}}
    \caption{Simulated lightcurves illustrating the mixed OU
      process. The lightcurves were generated using $\mu = 0, \varsigma =
      1, \tau_{H} = 0.1, \tau_{L} = 10$, independent Gaussian white
      driving noises, and a common random number generator seed. The
      three lightcurves probe the white noise (flat) part 
      of the power spectrum (top), the `pink' noise ($1 / f$) part of the
      power spectrum (middle), and the red noise ($1 / f^2$) part of
      the power spectrum. The mixed OU process can look very different
      depending on the time scales probed by the observations.
    \label{f-mixou_illust}}
  \end{center}
\end{figure}

  In Figure \ref{f-mixou_illust} we plot a mixed OU process sampled at
  the three different time intervals. The mixing weights for this
  example were chosen such that the power spectrum of the simulated
  lightcurve was flat below a low-frequency break, $\omega_L$, decays
  as $1 / f$ above the low-frequency break, and then steepened to $1 /
  f^2$ above a high-frequency break, $\omega_H$, similar to what is
  seen in the X-ray lightcurves of accreting black hole systems. Thus,
  the simulated lightcurves probe the `white' noise, `pink' 
  noise, and `red' noise regions of the PSD. We constructed
  these lightcurves assuming that the logarithm of the flux follows a
  mixed OU process with $\mu = 0, \varsigma = 1, \tau_{H} = 0.1 = 1 / \omega_H,$ and
  $\tau_{L} = 10 = 1 / \omega_L$, and that the driving noise is Gaussian white
  noise. In general, we model the logarithm of the flux as a Gaussian mixed-OU process, as this is consistent with the flux-rms correlation seen in the X-ray variations of GBHs and AGN \citep{uttley01} and the log-normal distribution of fluxes for Cygnus X-1 \citep{uttley05b}. In addition, the same seed for the random number simulator 
  was used in order enable a more direct comparison between the
  lightcurves. As is apparent, the mixed OU process can appear very
  different depending on which time scales are probed by the observation.

  Equation (\ref{eq-Pmixou_asympt}) in the Appendix demonstrates that the power
  spectrum of the continuous mixed OU process, with mixing function given by
  Equation (\ref{eq-mixfunc}), exhibits the same behavior that is seen
  in the power spectrum of the X-ray lightcurves of GBHs and AGN. This
  therefore suggests that the mixed OU process well represents the
  fluctuations in the X-ray lightcurves of GBH and AGN, and may hold
  clues to the physical origin of the X-ray fluctuations. Moreover, we
  have identified a process which gives an explicit 
  connection between the break frequencies in the X-ray power spectra
  and characteristic time scales. This is important, because it is
  not always apparent how to connect a feature in the power spectrum
  with behavior in the time-domain. For example, we note that the
  characteristic time scales correspond to break frequencies when
  calculating the power spectrum in terms of $\omega$, but simply
  taking the inverse of the ordinary frequency, $1 / f$, corresponding
  to the break will overestimate $\tau_{L}$ and $\tau_{H}$ by a
  factor of $2\pi$.

  Physically, a mixed OU process may represent the X-ray fluctuations
  of GBH and AGN under the propagating disturbance model, originally
  developed by \citet{lyub97}. In this model, stochastic disturbances
  originate in the region of the accretion disk outside of the X-ray
  emitting region. These disturbances propagate inwards and modulate
  the X-ray emitting region, which cause fluctuations in the X-ray
  flux.  Because many physically relevant time scales increase with
  radius, the X-ray flux varies on a broad range of time scales due to
  the disturbances originating over a broad range in radii. If the
  disturbances at various radii are independent and are well described
  by an OU process with characteristic time scales that increase with
  radius, then the X-ray flux should be well described by a mixture of
  OU processes. The fact that we can construct a mixed OU process with
  PSD similar to the X-ray power spectra of GBH and AGN 
  is consistent with this interpretation. In this case, the break frequencies
  correspond to the characteristic time scales of the disturbances at
  the maximum and minimum radii of the region that contributes to modulating the
  X-ray emission. We develop this interpretation further in
  \S~\ref{s-oudiff}.

\section{Mixed OU Process as Resulting from Stochastic Diffusion in the
  Accretion Disk}

\label{s-oudiff}

Similar to us, many authors have attempted to identify a particular
stochastic process for the accretion rate perturbations. While we
model the accretion rate perturbations as following an OU process, \citet{arev06b},
\citet{tit07}, and \citet{misra08} have shown that a power spectrum
with a $1 / f$ region can also be created when the accretion rate
perturbations have a Lorentzian power spectrum, i.e., when the
disturbances are quasi-periodic. Perturbations of this sort are
expected if, for example, the perturbations are due to Rayleigh-Taylor
instability \citep{tit07}. \citet[][see also \citet{mayer06} and
\citet{janiuk07}]{king04} 
model the fluctuations as resulting from a magnetic field modeled as a
first order autoregressive process, which is a discrete form of the OU
process \citep[e.g.,][]{brockwell02}. We can build on the work of
\citet{wood01} and \citet{tit07} to show that a mixed OU process can
model the viscous response of the 
accretion disk subject to a stochastic driving noise field, for the
special case when the diffusion equation is linear.  

Conservation of mass and angular momentum result in material in the
disk obeying a nonlinear diffusion equation: 
\begin{equation}
  \frac{\partial \Sigma}{\partial t} = \frac{3}{r} \frac{\partial}{\partial r}\left[ r^{1/2}
    \frac{\partial}{\partial r} \left(r^{1/2} \nu \Sigma \right)
  \right]. 
  \label{eq-nonlinear_diff}
\end{equation}
Here, $\Sigma$ is the surface density of the disk and $\nu$ is the
kinematic viscosity. Under the assumption that the rotational
frequency of the disk is equal to the Keplerian value, the accretion
rate is given by \citep{wood01} 
\begin{equation}
  \dot{M}(r,t) = 6 \pi r^{1/2} \frac{\partial}{\partial r} (r^{1/2} \nu \Sigma).
  \label{eq-mdot}
\end{equation}
We can use Equations (\ref{eq-nonlinear_diff}) and (\ref{eq-mdot}) to
study the evolution of accretion rate perturbations about the average
steady state solution. Following \citet{wood01} and \citet{tit07},
define $x = r^{1/2}$ and a perturbation to the quantity $r^{1/2} \nu
\Sigma$ as $u(x,t) = x \Delta(\nu \Sigma)$, and therefore $\Delta
\dot{M}(r,t) \propto \partial u(x,t) / \partial x$. The nonlinear
diffusion equation for a single perturbation can now be written as
\citep[e.g.,][]{frank02}
\begin{equation}
  \frac{\partial u}{\partial t} = \frac{3}{4} \frac{\partial
    u}{\partial \Sigma} \frac{\partial^2 u}{\partial x^2}.
  \label{eq-nonlinear_perturb}
\end{equation}
When $\nu$ is independent of $\Sigma$, Equation
(\ref{eq-nonlinear_perturb}) becomes a linear diffusion equation. 

Equation (\ref{eq-nonlinear_perturb}) only describes the evolution of
a single perturbation, but in order to create flickering in $u(x,t)$
we need to introduce a persistent source of the fluctuations. We can introduce
fluctuations to $u(x,t)$, and therefore to $\Delta 
\dot{M}(r,t)$, by introducing a stochastic source term to Equation
(\ref{eq-nonlinear_perturb}).  For simplicity, we only study the
linear form of Equation (\ref{eq-nonlinear_perturb}), and therefore
assume that the viscosity is independent of surface density. While it
is unlikely that the viscosity is independent of $\Sigma$, it is still
illuminating to study the linear form of the diffusion equation in order
to better understand how the disk responds to a stochastic driving
noise, as it permits an analytical treatment. Denote the stochastic
source as $\partial W(x,t) / \partial t$, where $\partial W(x,t)
/ \partial t$ is a spatially dependent
white noise \footnote[2]{This notation should not be strictly
  interpreted as a traditional derivative, as the definitions and
  rules used in stochastic calculus are not necessarily the same as
  those of ordinary calculus. However, we ignore this mathematical
  technicality, as the notation $\partial W(x,t) / \partial t$ conveys
  the appropriate interpretation of $W(x,t)$ as the integral of
  $\partial W(x,t) / \partial t$ over $t$.}, and therefore $W(x,t)$ is a Wiener random field. By
definition, $\partial W(x,t) / \partial t$ has a flat power
spectrum and zero mean. The mean of $W(x,t)$ is also zero, and the
covariance of $W(x,t)$ is 
\begin{equation}
  E\left[ W(x,t) W(y,s) \right] = \varsigma(x,y) \min(t,s),
  \label{eq-wiener_covar}
\end{equation}
where $E(x)$ denotes the expected value of $x$, and $\varsigma(x,y)$
defines the spatial covariance of the Wiener random field. The Wiener
process is essentially a continuous time Brownian motion, and
therefore the Wiener random field $W(x,t)$ may be viewed as an
infinite set of Brownian motions indexed by the parameter $x$. If
there are no spatial correlations for $W(x,t)$, then $\varsigma(x,y) =
\delta(x-y)$ and $\partial W(x,t) / \partial t$ is called a
`space-time' white noise; however, in this case, solutions to the
linear stochastic diffusion equation only exist for the case of one
spatial dimension \citep{chow07}. 

We assume that the perturbations to the accretion rate only exist in a bounded accretion flow defined
over a region $0 < x < x_{max}$, where $x_{max}$ represents the outer
edge of the flow.  We adopt the boundary condition given by 
\citet{wood01} and \citet{tit07}, with $u(0,t) = 0$ and $\partial
u(x_{max},t) / \partial t = 0$. The first boundary condition assures
that the perturbations of $\Sigma$ go to zero at the inner boundary of
the disk, and the second ensures the the disk accretes the
perturbations, i.e., the perturbations can only leave through the
inner boundary. Including a stochastic source in the linear diffusion
equation, our stochastic model for the accretion rate perturbations is
defined by the following set of Equations: 
\begin{eqnarray}
  \frac{\partial u(x,t)}{\partial t} & = & \frac{3\nu(x)}{4}
  \frac{\partial^2 u(x,t)}{\partial x^2} + \frac{\partial
    W(x,t)}{\partial t} \label{eq-linear_perturb} \\ 
  \Delta \dot{M}(x,t) & = & 3\pi \frac{\partial u(x,t)}{\partial
    x} \label{eq-mdot_perturb} \\ 
  u(x,0) & = & u_0(x) \label{eq-icondition} \\
  u(0,t) & = & 0 = \frac{\partial u(x_{max},t)}{\partial t}. \label{eq-bc}
\end{eqnarray}
Equations (\ref{eq-linear_perturb})--(\ref{eq-bc}) can be solved using
standard methods developed for stochastic partial differential
equations, and a good introduction is \citet{chow07}. 

Before presenting the solution, a couple of remarks are
warranted. First, the model defined by Equations
(\ref{eq-linear_perturb})--(\ref{eq-bc}) is clearly only an approximation for
several reasons. As mentioned before, the actual viscous response of
the disk will be nonlinear, as the viscosity also likely 
depends on the surface density. In addition, for the noise process
assumed above, there is nothing to prevent the linear model from
producing negative values of $\Sigma$ and $\dot{M}$, which are
unphysical. A better approach would be to postulate a stochastic model
for the viscosity which ensures positivity of the solution, and then solve Equation
(\ref{eq-nonlinear_perturb}) numerically using this perscription for
$\nu(r,\Sigma,t)$. However, in this work our goal is to study the
simpler model, for which analytical solutions are possible, in order
to better understand how the disk responds to a random driving noise
field, and thus provide some insight on how to connect features in the power
spectra to the physics of the accretion disk. Moreover, we will show
that the solution to the linear diffusion equation is a mixture of
OU-processes, providing justification for their use in modeling
variability. Second, it is important to note that the stochastic noise
field in the partial differential equation, $\partial W(x,t)
/ \partial t$, is the time derivative of the process which is the
source of the random perturbations, and is not the process itself. And
third, the physical interpretation of the noise process is left
unspecified, however, as many authors have suggested, $\partial
W(x,t) / \partial t$ may represent fluctuations in $\partial u(x,t)
/ \partial t$ which are the result of turbulent and chaotic MHD
effects \citep[e.g.,][]{lyub97,king04}. Indeed, MHD turbulence has
been observed in numerical simulations \citep[e.g.,][]{hawley01}. 

The solution to Equations (\ref{eq-linear_perturb})--(\ref{eq-bc}) can
be expressed through the eigenfunctions of the diffusion operator. The
eigenfunctions, $e_k(x)$, and eigenvalues, $\omega_k$, are defined by 
\begin{equation}
  \frac{3\nu(x)}{4x^2} \frac{d^2}{dx^2} e_k(x) = \omega_k e_k(x), \label{eq-eigensystem}
\end{equation}
subject to the appropriate boundary conditions. Eigenfunctions and
eigenvalues are given in \citet{wood01} and \citet{tit07},
respectively, for the case where the viscosity $\nu(x)$ is a power-law
in $x$. The eigenfunctions are orthonormal with respect to the
weighting function $p(x) = 4 x^2 / (3\nu(x))$, and form a complete
set. The solution to Equations
(\ref{eq-linear_perturb})--(\ref{eq-bc}) is then given by
\citet{chow07} 
\begin{equation}
  u(x,t) = \sum_{k=1}^{\infty} u_k(t) e_k(t) \label{eq-solution},
\end{equation}
where $u_k(t)$ are a set of stochastic processes. In order to
determine $u_k(t)$, we first note that if the spatial covariance
function of the driving noise, $\varsigma(x,y)$, exists in the space
spanned by the eigenfunctions, then we can express it as an
eigenfunction expansion, 
\begin{equation}
  \varsigma(x,y) = \sum_{k=1}^{\infty} \varsigma^2_k e_k(x) e_k(y),
  \label{eq-sigexpand}
\end{equation}
where the coefficients are
\begin{equation}
  \varsigma^2_k = \int_{0}^{x_{max}} \int_{0}^{x_{max}} \varsigma(x,y) p(x) e_k(x) p(y) e_k(y)\ dx\ dy.
  \label{eq-sigcoefs}
\end{equation}
The random field $W(x,t)$ can then be expressed as
\citep[e.g.,][]{chow07} 
\begin{equation}
  W(x,t) = \sum_{k=1}^{\infty} \varsigma_k e_k(x) w_k(t),
  \label{eq-wexpand}
\end{equation}
where $\{ w_k(t) \}$ are a sequence of independent and identically
distributed Brownian motions. 

Inserting Equations (\ref{eq-eigensystem}), (\ref{eq-solution}), and
(\ref{eq-wexpand}) into Equations (\ref{eq-linear_perturb}) and
(\ref{eq-icondition}) produces an infinite set of ordinary stochastic
differential equations: 
\begin{eqnarray}
  du_k(t) &  = & -\omega_k u_k(t) dt + \varsigma_k dw_k(t), \ \ \ k = 1, 2, \ldots, \label{eq-ukt} \\
  u_k(0) & = & \int_{0}^{x_{max}} u_0(x) p(x) e_k(x)\ dx \label{eq-uk0}.
\end{eqnarray}
Comparison of Equations (\ref{eq-ukt}) and (\ref{eq-uk0}) with
Equation (\ref{eq-ou}) shows that $u_k(t)$ are a set of independent OU
processes with initial condition $u_k(0)$. Therefore, the solution to
the linear diffusion equation defined by Equations
(\ref{eq-linear_perturb})--(\ref{eq-bc}) is  
\begin{eqnarray}
  u(x,t) & = & \sum_{k=1}^{\infty} \left[ u_k(0) e_k(x) e^{-\omega_k t} + \varsigma_k e_k(x) X_{ou}(t,\omega_k) \right]
  \label{eq-linsolution} \\
  \Delta\dot{M}(x,t) & \propto & \sum_{k=1}^{\infty} \left[ u_k(0)
    \frac{\partial e_k(x)}{\partial x} e^{-\omega_k t} + \varsigma_k
    \frac{\partial e_k(x)}{\partial x} X_{ou}(t,\omega_k) \right] 
\end{eqnarray}
where $X_{ou}(t,\omega_k)$ is an OU process defined by Equation
(\ref{eq-ou}) with white driving noise, $\mu = 0$, and $\varsigma =
1$. The first eigenvalue is roughly given by the viscous timescale at
the outer edge of the disk, $\tau_{\rm visc}$, depending on how
$\nu(x)$ depends on $x$. In particular, if $\nu(x)$ has a power-law
dependency on $x$, then $\omega_1 \propto 1 / \tau_{\rm visc}$
\citep{tit07}.  In addition, if the driving noise field is a Gaussian
process, then the accretion rate perturbations will also be a Gaussian
process.

In general, we are interested in the stationary solution, which occurs
when $t \rightarrow \infty$. As a practical matter, this occurs after
a few viscous times have passed. In this case, the initial
condition has been forgotten, and the solution to the linear
stochastic diffusion equation is a mixture of OU processes, with the characteristic frequencies given by the eigenvalues of the linear diffusion operator, the
mixing weights given by the product of the eigenfunctions of the linear diffusion
operator and the coefficients of the eigenfunction expansion of the spatial covariance of the driving noise field. Then, the covariance function of the
accretion rate perturbations is 
\begin{equation}
Cov(\Delta \dot{M}(x,t), \Delta \dot{M}(y,s)) = \sum_{k=1}^{k=\infty}
\frac{\varsigma_k^2}{2\omega_k} \frac{\partial e_k(x)}{\partial x}
\frac{\partial e_k(y)}{\partial y} e^{-\omega_k |t-s|},
\label{eq-mdot_covar}
\end{equation}
and the power spectral density for fluctuations as a function of
radius is 
\begin{equation}
  P_{\Delta \dot{M}}(x, \omega) = \sum_{k=1}^{\infty}
  \frac{\varsigma_k^2}{2\pi} \left( \frac{\partial e_k(x)}{\partial x}
  \right)^2 \frac{1}{\omega_k^2 + \omega^2}.
  \label{eq-mdot_pspec}
\end{equation}

The X-ray emission of accreting black holes is thought to be released in
the inner regions of the disk, and therefore the behavior of the
accretion rate fluctuations at small $x$ is of primary interest.  It
is illuminating to investigate asymptotic behavior of the power
spectrum at $x \rightarrow 0$ for the case where $\nu(x) \propto
x^{\alpha}$. As per the discussion in \S~\ref{s-supou}, the power
spectrum of a mixed OU process is flat on frequencies shorter than $\omega_1$,
and therefore the fluctuations are decorrelated on time scales longer
than the viscous time scale at $x_{max}$. However, on time scales
short compared to the viscous time scale, the situation is more
complicated as the power spectrum also depends on the spatial
covariance structure of the driving noise field, $\varsigma(x,y)$,
through the set of $\varsigma_k$. For simplicity, we first assume that
there is no spatial correlation in the driving noises, and therefore
$\varsigma(x,y) = \delta(x-y)$ and $\varsigma_k = 1$ for all $k$. Following
\citet{tit07}, one can then show that the power spectrum of the
accretion rate perturbations at small radii decays as a power-law on
timescales short compared to the viscous time scale, with the slope of
the power-law decay steepening when the viscosity increases less
steeply toward larger raddi. Therefore, if the viscosity increases more
steeply toward higher radii, then at small radii the longer time scale
perturbations are more suppressed relative to the short time scale
perturbations.

In order to investigate the effect that the spatial correlations in
the noise field have on the observed power spectra of $\Delta
\dot{M}(x=0,t)$, we study the case of $\nu(x) \propto x^2$. In this case the
eigenfunctions are $\sin$ functions and the eigenvalues are $\omega_k
\propto (2k - 1)^2$ \citep[e.g.,][]{wood01}. One can show that the
power spectrum for $\Delta \dot{M}$ at small radii is then
\begin{eqnarray}
  P_{\Delta \dot{M}}(x, \omega) \propto \sum_{k=1}^{\infty}
  \frac{\varsigma_k^2 \omega_k}{\omega_k^2 + \omega^2} \label{eq-mdot_pspec2} \\
  \omega_k = (2k - 1)^2 \frac{\pi}{2}.\label{eq-evalues}
\end{eqnarray}
As an example, suppose that the spatial correlations for $W(x,t)$
decay exponentially in radius with some characteristic radius $r_0$: 
\begin{equation}
  \varsigma(x,y) \propto e^{-|x^2 - y^2| / r_0} \label{eq-sigxy_example}.
\end{equation}
Then, in this case, the coefficients in the eigenfunction expansion,
$\varsigma_k^2$, are constant for $\omega_k \ll r_0 / r_{max}$ and fall
off as $\varsigma_k^2 \propto 1 / \omega_k^2$ for $\omega_k \gg r_0 /
r_{max}$. Inserting this form for $\varsigma_k^2$ into the Equation for
the power spectrum of $\Delta \dot{M}(0,t)$ suppresses the higher
frequency modes in the eigenfunction expansion, thus supressing the
fluctuations on short time scales. In particular, for a power spectrum of
the form Equation (\ref{eq-mdot_pspec2}), a second break will appear
near $\omega_{r_0} \sim 1 / \tau_{r_0}$, producing a second region which
decays even faster than the intermediate region. The second
characteristic time scale, $\tau_{r_0}$, corresponds to the time it
takes a perturbation traveling at the viscous speed, $v_r \sim r_{max}
/ \tau_{\rm visc}$, to travel across a region corresponding to the
spatial scale of the driving noise field:
\begin{equation}
  \tau_{r_0} \sim \frac{r_0}{v_r} \sim \frac{r_0}{r_{max}}\tau_{\rm visc}
  \label{eq-tau0}
\end{equation}
This may be the source of the high frequency break seen in X-ray power
spectra of GBH and AGN, although it may also be due to an abrupt
change in the accretion geometry or viscosity \citep{churaz01,psaltis00}. Furthermore, we note
that if the amplitude of 
the driving noise field increases with radius, then $\varsigma_k$ will
decrease with increasing $k$, steepening the slopes in both the
intermediate region ($\omega_1 \lesssim \omega \lesssim \omega_{r_0}$)
and the high-frequency region $\omega \gtrsim \omega_{r_0}$. However, a decrease in viscosity with increasing radius also has this effect, and thus the two situations are degenerate. 

Finally, we note that in order to connect the power-spectrum of
luminosity fluctations with that of accretion rate fluctuations, heat
and photon diffusion effects must also be considered. The diffusion of
heat and photons are both governed by diffusion equations, and the
radiative output of the disk may be described as resulting from the
convolution of the solution to the heat and photon diffusion equations
with the accretion rate fluctuations \citep[e.g.,][]{tit07}, which are
stochastic. The power spectrum of the radiative output is then the
product of the power spectra for the accretion rate perturbations and
the power spectra for the solution to the heat and photon diffusion
equations.  While a detailed treatment of the effects of heat and
photon diffusion is beyond the scope of this work \citep[but
see][]{sun80,sun85,tit07}, we note that based on the above discussion
and the results of \citet{tit07}, the power spectra of the solution to
the linear heat and photon diffusion equations will also be of the
form of Equation (\ref{eq-Pmixou_disc}).  Therefore, additional breaks
will occur at frequencies corresponding to the characteristic
frequencies of the heat and photon diffusion equations, and the power
spectra of the radiative output will steepen even further above these
breaks. Thus, the observed fluctuations in the emission from accreting
black holes may be interpreted as the result of the response of the
disk to a random noise field, where the viscous, thermal, and
radiative response of the disk act as a series of low-pass filters on
the driving noise field, sequentially suppressing the variability on
time scales short compared to the characteristic time scales for
viscous, thermal, and photon diffusion. Furthermore, if the noise
field is spatially correlated, then variability on time scales short
compared to the drift time scale for the characteristic length of
the spatial correlations is also suppressed. Thus, breaks are expected
to exist in the power spectra at the frequencies corresponding to
these time scales, and the slope of the power spectra in the different
regions contains information on the viscous, thermal, and radiative
structure of the accretion flow, as well as the structure of the
driving noise field. 

For completeness, we summarize here the some of the results from the above discussion:
\begin{itemize}
\item The mixed OU process is the solution to the linear diffusion
  equation, subject to a random driving noise field. The mixing
  weights are the product of the eigenfunctions of the linear
  diffusion operator and the coefficients of the eigenfunction
  expansion of the covariance function of the driving noise field. The
  characteristic frequencies of the OU processes are the eigenvalues
  of the linear diffusion operator, subject to the appropriate
  boundary conditions. 
  \item The power spectrum of accretion rate perturbations at small
    radii, where most of the energy is thought to be released, is
    flat on time scales longer than the viscous time scale at the
    largest radii. On timescales shorter than $\tau_{\rm visc}$, the
    power spectrum has the form $C_1  / \omega^2 < P_{\Delta
      \dot{M}}(\omega) < C_2$, where $C_1$ and $C_2$ are constants. 
  \item There is a second break in the power spectrum of $\Delta
    \dot{M}$ on time scales short compared to the crossing time of a
    perturbation traveling at the viscous speed across a region of
    length similar to the characteristic spatial correlation length of
    the driving noise field. The power spectrum of $\Delta \dot{M}$
    steepens on frequencies higher than this break. 
  \item If the viscosity of the disk increases toward higher radii,
    then in the inner region of the disk the long time scale
    fluctuations are surpressed compared to the short time scale
    fluctuations, thus flattening the power spectrum. Similarly, if
    the amplitude of the driving noise increases toward smaller 
    radii, the power spectrum also flattens. Thus, these two effects
    are degenerate and cannot be distinguished on the basis of power
    spectra alone. 
  \item Heat and photon diffusion can introduce additional breaks in
    the power spectra of the emission from accreting black holes, with
    the power spectra steepening above these breaks.  
\end{itemize}

  \section{Fitting the Mixed OU Process}

  \label{s-fits}

  In the previous section we have suggested an interpretation of the mixed OU
process developed in this work in terms of the viscous response of the
accretion disk subject to a random driving noise field. However, the
mixed OU process is not limited to linear diffusion models for viscous
evolution of the accretion disk, and in 
fact one can go through the same steps outlined above to show that the
mixed OU process is a solution for many parabolic stochastic partial
differential equations, and thus will also describe, say, the
diffusion of heat in the disk, which is governed by the heat
equation. However, regardless of the physical mechanism, the mixed OU
process provides an accurate statistical 
description of the lightcurves of GBH and AGN, in the sense that
it reproduces much of the shape of the power spectra of these
objects. As a result, estimation of the break frequencies can be done
directly from the observed lightcurve, and is therefore 
free of the biases due to windowing effects, which are inherent in
fitting of power spectra. Some features are not captured by this
process, such as quasi-periodic oscillations and nonlinear effects. An additional quasi-periodic
component can be modeled as a solution to a set of hyperbolic
stochastic partial differential equations, in particular the random
wave equation. However, inclusion of quasi-periodic oscillations to
our model is beyond the scope of this work. 

\subsection{The Likelihood Function}

\label{s-likhood}

  The mixed OU process may be fit using maximum-likelihood or Bayesian
  techniques. This has
  the advantage that the parameters for the process are estimated
  directly from the observed lightcurve utilizing all of the
  information in the data. In particular, the estimates of the
  parameters for the power spectrum model are not biased
  by measurement errors, irregular sampling, or other windowing effects
  due to the finite time span of the lightcurve, such as red noise
  leak. While these biases may be corrected for using Monte Carlo
  methods \citep{uttley02}, such methods are also computationally
  intensive and rely on a $\chi^2$ statistic. Indeed, these
  benefits of time domain modeling are our primary motivation in 
  developing the mixed OU process model; we sought to obtain a means
  of estimating the break frequencies and variability amplitude
  of X-ray lightcurves in a manner that was free of the biases that
  can affect frequency domain techniques.

  In order to calculate the likelihood function of the lightcurve, it
  is necessary to assume a probability distribution for the driving
  noise process. In this work we assume that the driving noise $dW(t)$
  is Gaussian white noise (i.e., $W(t)$ is the Gaussian Wiener process). In the
  context of the linear diffusion equation interpretation, this is
  equivalent to assuming that the driving noise field is a Gaussian
  process. Denote the 
  $n$ observed flux values as $y_1,\ldots,y_N$, sampled at time values
  of $t_1,\ldots, t_N$. The observed flux values have independent
  Gaussian measurement errors with variances $v_1,\ldots,v_N$. We denote
  ${\bf y,t,v}$ to be the vectors containing the $N$ values of flux,
  time, and measurement error variance. In this case it is
  straightforward to calculate the likelihood function, as the
  lightcurves follow a $N$-dimensional multivariate normal distribution
  with mean $\mu$ and covariance matrix $R_{Y}({\bf t}) + {\rm diag}({\bf v})$, where
  $R_{Y}({\bf t})$ is given by Equation (\ref{eq-Rmixou_disc}) calculated at
  the appropriate values, and ${\rm diag}({\bf v})$ operates on the
  vector ${\bf v}$ by constructing a diagonal matrix with ${\bf v}$ along the
  diagonal.

  Unfortunately, direct calculation of the
  likelihood becomes computationally prohibitive for most reasonable
  values of $N$. However, for the discrete mixture the likelihood for an observed
  lightcurve, modeled according to Equation (\ref{eq-mixou_disc}),
  can be efficiently calculated by 
  obtaining a state-space representation of the lightcurve, and then
  using the Kalman recursions \footnote[1]{State-space representations
    and the Kalman recursions are commonly used in time series
    analysis. A good introduction to these techniques is given by
    \citet{brockwell02}.}. Using these techniques, the likelihood
  function, $p({\bf y}|{\bf c},{\bf \omega},\mu,\hat{\varsigma})$,
  is calculated using the following recursion formulae:
  \begin{eqnarray}
    p({\bf y}|{\bf c},{\bf \omega},\mu,\hat{\varsigma}) & = &
    \prod_{i=1}^n \left[2\pi \Delta_i\right]^{-1/2}
    \exp\left\{ -\frac{1}{2} \frac{(\hat{y}_i - y_i)^2}{\Delta_i} \right\}
    \label{eq-lik} \\
    \hat{y}_1 & = & \mu \\
    \Delta_1 & = & \frac{\varsigma^2}{2} \sum_{j=1}^M c_j^2 \tau_j \\
    \hat{\bf x}_1 & = & {\bf 0} \label{eq-xhat0} \\
    {\bf \Omega}_1 & = & {\rm diag}\left(\frac{\varsigma^2}
      {2}\left[\frac{c_1}{\omega_{1}},\ldots,\frac{c_M}{\omega_{M}}\right]^T\right) \label{eq-xhvar0} \\
    {\bf A}_i & = & {\rm
      diag}\left(\exp\{-(t_i-t_{i-1})[\omega_{1},\ldots,\omega_{M}]^T
      \}\right) \label{eq-Amat} \\
    {\bf \Theta}_i & = & {\bf A}_i {\bf \Omega}_i {\bf c} \\
    \hat{\bf x}_{i} & = & {\bf A}_i \hat{\bf x}_{i-1} + \frac{\bf
      \Theta_{i}}{\Delta_{i-1}} (y_{i-1} - \mu - {\bf c}^T \hat{\bf
      x}_i) \label{eq-xhati} \\
    {\bf \Omega}_{i} & = & {\bf A}_i {\bf \Omega}_{i-1} {\bf A}_i^T +
    {\rm diag}({\bf u}_{OU,i}) - {\bf \Theta}_i {\bf \Theta}_i^T /
    \Delta_i \label{eq-omegai} \\
    u_{OU,ij} & = & \frac{\varsigma^2}{2\omega_{j}} \left(1 -
      e^{-(t_i-t_{i-1})\omega_{j}}\right) \label{eq-voui} \\
    \hat{y}_i & = & \mu + {\bf c}^T \hat{\bf x}_i \label{eq-yhati} \\
    \Delta_i & = & {\bf c}^T {\bf \Omega}_i {\bf c} + v_i. \label{eq-deltai}
  \end{eqnarray}
  Here, ${\bf c}$ and ${\bf \omega}$ denote the $M$-dimensional
  vectors of mixing weights and characteristic frequencies,
  respectively, ${\bf x}^T$ denotes the transpose of ${\bf x}$, and
  Equations 
  (\ref{eq-Amat})--(\ref{eq-deltai}) are only valid for $i >
  1$. Recall that the $M$ characteristic time scales are $\tau_1 = 1 /
  \omega_1, \ldots, \tau_M = 1 / \omega_M$. Equation (\ref{eq-lik})
  can then be maximized to calculate 
  maximum-likelihood estimates of ${\bf c}, \mu,{\bf \omega},$ and
  $\varsigma$, or used to perform Bayesian inference in combination with a
  prior distribution. Precise determination of the mixing weights will in
  general only be possible with the highest quality
  lightcurves, as the estimates will have a large uncertainty with high
  levels of correlation in their error distribution. 

  In practice, an additional constraint
  needs to be imposed on the norm of ${\bf c}$, as it is
  degenerate with $\varsigma$. In this work we impose a unit squared norm constraint on the weights:
  \begin{equation}
    ||{\bf c}||^2 = \sum_{i=1}^M c^2_i = 1.
    \label{eq-unitnorm}
  \end{equation}
  Under this normalization, the PSD at frequencies $\omega \gg {\rm max}\{\omega_1,\ldots,\omega_M\}$ is
  \begin{equation}
    P(\omega) \sim \frac{\varsigma^2}{2\pi \omega^2},
    \label{eq-highf_psd}
  \end{equation}
  from which it is apparent that the amplitude of the high frequency
  PSD is proportional to $\varsigma^2$. Thus, $\varsigma$ not only gives rate at which variability power is injected into the stochastic process, but also
  defines the amplitude of variability at the highest frequencies. The amount of high frequency variability in a time interval $t_1 < t < t_2$ is given by the integral of Equation (\ref{eq-highf_psd}) over this time interval:
  \begin{equation}
    Var(t_1 < t < t_2) = \frac{\varsigma^2}{2\pi} \left( t_2 - t_1 \right), \ \ \ t_2 << 1 / \omega_H.
    \label{eq-highf_var}
  \end{equation}

  So long as $M \ll N$, calculation of the likelihood via Equations
(\ref{eq-lik})--(\ref{eq-deltai}) is much faster than direct
calculation using the autocovariance matrix, $R_{Y}({\bf t})$.  In
addition, we note that Equation (\ref{eq-lik}) is only valid for
Gaussian measurement errors, and therefore, one must have obtained a
suitably high number of counts in order to use it for X-ray and gamma ray data. An
extension to the low-count Poisson regime is beyond the scope of our
current paper, but will be developed in future work (Kelly et al., in
preparation).

  Because we are primarily interested in bending power-law forms of
  the power spectrum, we use a mixed OU process which can accurately
  model these forms. We have found that the following weighting scheme
  on a regular grid in $\log \omega_{j}$ can provide a good
  approximation to a bending power-law with two breaks:
  \begin{eqnarray}
   c_j & = & \omega_{j}^{1 - \alpha / 2} \left(\sum_{j=1}^M \omega_j^{2-\alpha}\right)^{-1/2} \label{eq-plaw_weights} \\
    \log \omega_{j} & = & \log \omega_L + \frac{j-1}{M - 1} \left( \log
      \omega_H - \log \omega_L \right), \ \ \ j = 1, \ldots, M
    \label{eq-omgrid}.
  \end{eqnarray}
  Here, the parameters of the process are $\mu, \varsigma, \omega_L,
  \omega_H,$ and $\alpha$. The mean of the process is $\mu$, $\varsigma$
  is the standard deviation of the driving noise, $\omega_H$ and
  $\omega_L$ are the high and low frequency breaks in the power
  spectrum, and $\alpha$ is the slope of the power spectrum in the
  intermediate region. The power spectrum for the process defined by
  Equations (\ref{eq-plaw_weights})--(\ref{eq-omgrid}) is flat for
  frequencies $\omega \lesssim \omega_L$, decays as $P(\omega) \propto
  1 / \omega^\alpha$ for $\omega_L \lesssim \omega \lesssim \omega_H$,
  and decays as $P(\omega) \propto 1 / \omega^2$ for $\omega \gtrsim
  \omega_H$. We have experimented with different values of
  $M$, and find that values of $M \gtrsim 30$ are more than
  sufficient; there is little change in the power spectrum among
  values of $M \gtrsim 30$.

  The likelihood function may be used to calculate a
  maximum-likelihood estimate. However, in order to get reliable
  estimates of the uncertainties on the mixed OU process parameters,
  we employ a Bayesian approach which calculates the posterior
  probability distribution of
  the parameters, given the observed lightcurve. The probability
  distribution of the parameters given the observed lightcurve is
  \begin{equation}
    p({\bf c}, {\bf \omega}, \mu, \varsigma|{\bf y})
    \propto p({\bf y}|{\bf c}, {\bf \omega}, \mu,
    \varsigma) p({\bf c}, {\bf \omega}, \mu, \varsigma),
    \label{eq-post}
  \end{equation}
  where the prior distribution is $p({\bf c}, {\bf \omega}, \mu,
  \varsigma)$. For the prior under the power-law weighting scheme, we fix
  the values of ${\bf c}$ and ${\bf \omega}$ to those in Equations
  (\ref{eq-plaw_weights}) and (\ref{eq-omgrid}), and assume a 
  uniform prior on $-2 < \alpha < 0$, $\mu$, and $\varsigma >
  0$. We assume the following prior on the logarithm of the break
  frequencies:
  \begin{eqnarray}
    p(\log \omega_L, \log \omega_H) & = & p(\log \omega_H|\log
    \omega_L) p(\log \omega_L) \nonumber \\
    & = & \frac{1}{\log \omega_{max} - \log \omega_L}, \ \ \
    \omega_{min} \le \omega_L \le \omega_H \le
    \omega_{max}. \label{eq-omprior}
  \end{eqnarray}
  Equation (\ref{eq-omprior}) is equivalent to placing a uniform prior
  on $\log \omega_L$ over the range $\omega_{min} \le \omega_{L}
  \le \omega_{max}$ and a uniform prior on $\log \omega_{H}$ over
  the range $\omega_{L} \le \omega_{H} \le \omega_{max}$. The
  upper and lower limits on the characteristic time scales that we search for, $\tau_{max} = 1 / \omega_{min}$ and $\tau_{min} = 1 / \omega_{max}$, are chosen to be 10 times the length of the time series, and $1 / 10$th
  the smallest time spacing in the time series. We use a Markov chain
  Monte Carlo (MCMC) algorithm to obtain random draws of the parameters from Equation
  (\ref{eq-post}).

  \subsection{Assessing the Quality of the Fit}

  \label{s-goodfit}

  We suggest two ways of assessing the quality of the mixed OU process
  in fitting the observed lightcurve. First, one should analyze the
  residuals to make sure that there are no systematic trends with
  time, and to ensure that there are no significant deviations from
  normality. If there are systematic trends with time, this may imply
  the existence of nonlinear effects. Define the standardized
  residuals as 
  \begin{equation}
    \chi_i \equiv \frac{y_i - \hat{y}_i}{\Delta_i}.
    \label{eq-sresids}
  \end{equation}
  Under the assumptions used to derive the likelihood function, the
  sequence of $\chi_i$ should follow a zero mean Gaussian white noise
  process with unit variance; i.e., there should be no systematic
  trends in $\{\chi_i\}$ with time and the distribution of the set of
  $\chi_i$ should be a Gaussian distribution with zero mean and
  variance equal to one. In practice, these assumptions are rarely
  true, although often the Gaussian likelihood approximation is
  adequate for estimating features in the power spectrum, such as
  characteristic time scales and the logarithmic slope of the power
  spectrum, so long as there are not many large outliers.

  The second method for assessing the quality of the fit which should
  be employed is to assess how well the mixed OU process fits the
  observed periodogram. This is similar in spirit to the Monte Carlo
  method of \citet{uttley02}, and enables one to assess whether there
  are features on different time scales which the mixed OU process does
  not fit well, as such discrepancies may be difficult to analyze in the
  lightcurve residual plot. First, one should calculate the periodogram
  of the observed lightcurve. Then, one can simulate a lightcurve having
  the same time sampling as the observed lightcurve for each of the
  mixed OU process parameters obtained from the MCMC
  sampler. Alternatively, one could also just use the maximum-likelihood
  estimate, but simulating lightcurves for each of the MCMC random draws
  also incorporates our uncertainty in the model parameters. Then, for
  each simulated lightcurve one calculates a periodogram. These
  periodograms can then be compared with the periodogram for the actual
  observed lightcurve to assess whether there are any time scales where
  the mixed OU stochastic process provides a poor fit.

  \section{APPLICATION TO BLACK HOLE LIGHTCURVES}

  \label{s-apply}

  In order to illustrate our method, and to investigate its
  effectiveness, we applied it to three different data sets. The first
  is an X-ray lighturve of the galactic black hole Cygnus X-1 in the
  low/hard state. For this object, the data is of high enough quality to clearly
  see two breaks in the power spectrum. The second data set is the
  sample of local AGN with long-term RXTE data studied by
  \citet{sob09b}, supplemented with XMM lightcurves when
  available. This sample of objects is well-studied, and broken
  power-law models have been estimated and reported by many
  authors. Both of these data sets provide a good test for our
  statistical model, as it allows us to compare it with other
  well-established methods. The third data set consists of optical
  lightcurves of AGN from the MACHO survey and the AGN Watch sample,
  taken from \citet{kbs09}. \citet{kbs09} analyzed these objects
  using only a single OU process, which is not as flexible as the
  mixed OU process. We exclude the lightcurves from the \citet{giveon99}
  sample used by \citet{kbs09}, as they are not as well sampled as
  the MACHO and AGN Watch sample.

  \subsection{Application to Cygnus X-1}

  \label{s-gbh}

\begin{figure}
  \begin{center}
    \scalebox{0.8}{\rotatebox{90}{\plotone{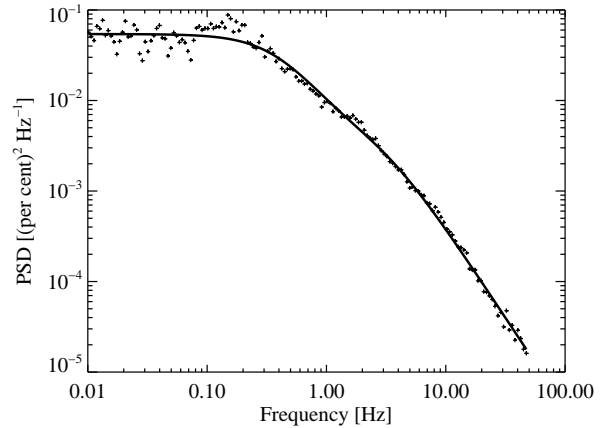}}}
    \caption{The observed PSD for an observation of Cygnus X-1 in the
      hard state (data points), compared with the best-fit PSD
      assuming the mixed OU process model (solid line). The mixed OU
      process model provides a good approximation to the observed
      PSD.
    \label{f-cygx1}}
  \end{center}
\end{figure}

  We used an archival RXTE observation of Cygnus X-1 from Oct 23, 1996
  (Obs ID 10241-01-01-000), which \citet{nowak99} have analyzed and identify as a hard-state
  lightcurve. We extracted the binned PCA lightcurve, using all channels,
  with the standard reduction routines. The lightcurve spanned $\sim
  28$ ksec and was binned in 0.01 sec intervals. The PSD for this
  observation is shown in Figure \ref{f-cygx1}, where we have
  subtracted off the Poisson noise level. We modeled the PSD
  using a mixture of 32 OU processes, treating the weights as free parameters on a
  fixed logarithmic grid in the characteristic frequencies, $\omega_j$; the
  maximum and minimum values of $\omega_j$ correspond to an order of
  magnitude shorter and longer than the minimum and maximum time
  spacing observed in the lightcurve. 

  Unlike for AGNs, the lightcurve has $\sim 1.5 \times 10^6$ data points,
  and computing the likelihood function is extremely computationally
  intensive. However, this large number of data points has the benefit
  that the PSD is very well determined. As a result, we do not
  estimate the parameters for the mixed OU process from the likelihood
  function of the lightcurve, but instead use least squares to
  fit the model PSD to the observed PSD. The result is also shown in
  Figure \ref{f-cygx1}. It should be noted that we do not bin the lightcurve as an
  integer number times the time-resolution of the instrument mode, and
  thus some aliasing is introduced to the periodogram. However, we do not consider this a
  concern as aliasing is likely to be negligible for the
  low-frequencies used in the fit. As can be seen, the mixed OU process is able
  to provide a good approximation to the PSD for this observation of
  Cygnus X-1, although it is not a perfect fit. Further improvement
  could be obtained by making the characteristic frequencies of the
  individual OU processes free parameters as well; however, we do not
  do this is this would increase the number of free parameters to
  $\approx 60$--$70$, making the least-square optimization
  computationally difficult. We consider this approach of directly
  fitting the PSD satisfactory for Cygnux X-1, even if it is not as
  powerful as maximum-likelihood, as 
  our primary goal in developing our method is to estimate features in
  the PSD, such as the break frequencies, from lightcurves that are of
  significantly poorer quality, for which high quality PSDs are not
  available. As such, we apply our model to Cygnus X-1 merely to
  illustrate that it provides a good approximation of the PSD.

  \subsection{AGN X-ray Lightcurves}

  \label{s-xrayagn}

  \subsubsection{Fitting Results}
  
  \label{s-xrayagn_fits}

  We applied our method to the X-ray lightcurves of 10 local AGN,
studied recently by \citet{sob09b}. These objects have high quality
lightcurves sampled over several decades in frequency, and have been
extensively studied using Fourier-based techniques, making them a good
sample for testing our method. The sample is summarized in Table
\ref{t-xray_agn}, which includes the values obtained previously in the literature from
Fourier-based techniques for the break frequency and logarithmic slope
of the power spectra above the break frequency. The 2-10 keV RXTE data
is presented in \citet{sob09b}, and is important for studying the
longer time scale behavior. In order to also study variations on time
scales $\lesssim 1$ day, we included XMM lightcurves from the
archives, when available. The XMM 2--10 keV background subtracted lightcurves were extracted
using the standard routines of the XMM Scientific Analysis System, and
were binned every 48 sec. Both the RXTE and XMM lightcurves were fit
simultaneously. The XMM 2--10 keV count rates were converted to 2--10
keV fluxes using the 2--10 keV flux values calculated by
\citet{demarco09}, with the exception of NGC 5506, where we used the
value calculated by \citet{guain10}. \citet{sob09b} calculated the 2--10 keV luminosities based on the best fit spectral model to the 3--20 keV PCA data.

  Our best fit parameters, as well as the 90\% confidence intervals,
  are also summarized in Table \ref{t-xray_agn}. In general, our values are
  consistent with values previously obtained by other authors using
  the Fourier-based Monte Carlo method of \citet{uttley02}, further
  validating our method. However, we obtain 
  consistently smaller errors on the estimated parameters. This is
  likely primarily due to the facts that previous 
  work treated the high frequency logarithmic slope of the power
  spectrum as a free parameter, while in our technique the high
  frequency end of the power law is assumed to decay as $1 /
  f^2$, and that we work directly with the likelihood function.

\begin{figure}
  \begin{center}
    \includegraphics[scale=0.33,angle=90]{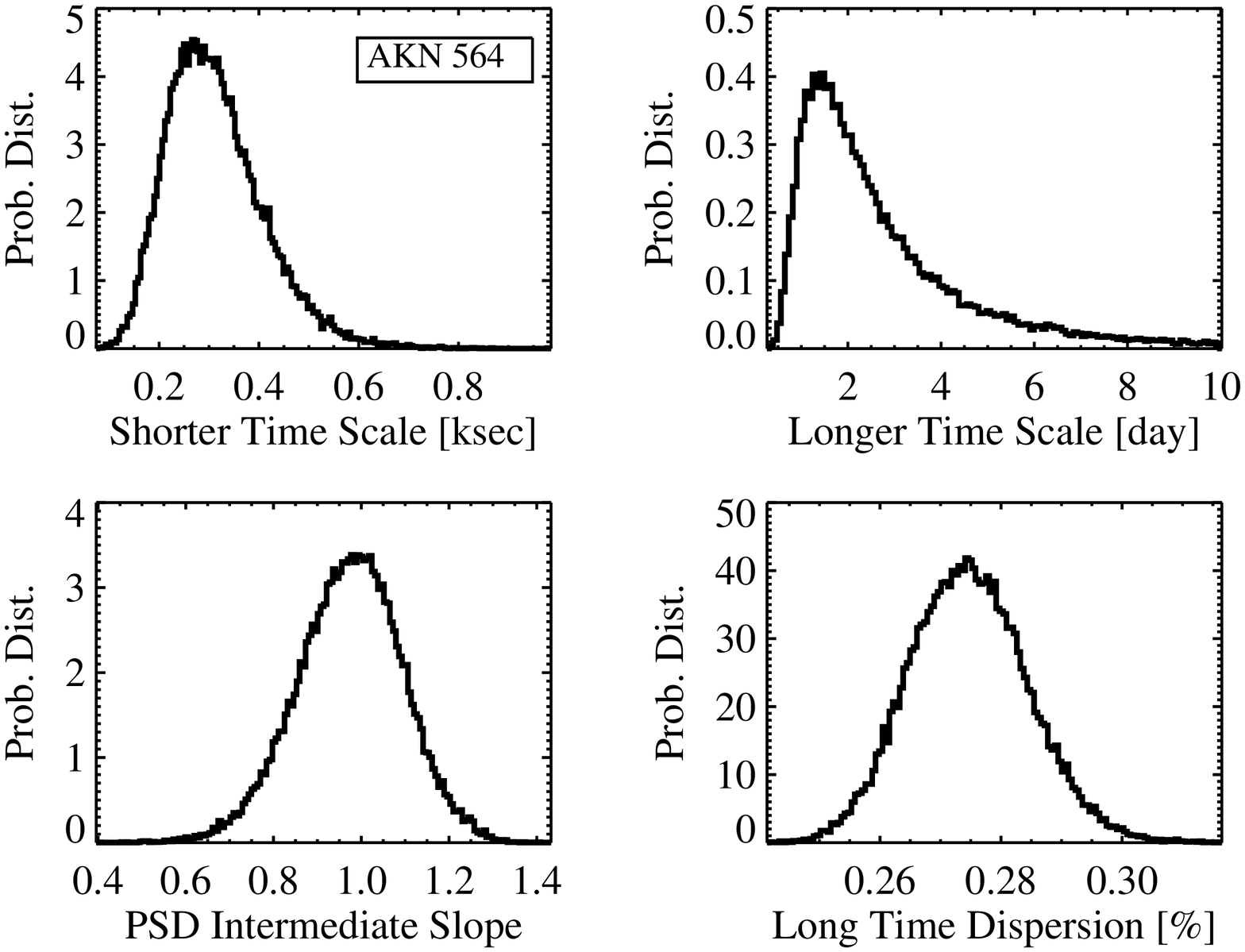}
    \includegraphics[scale=0.33,angle=90]{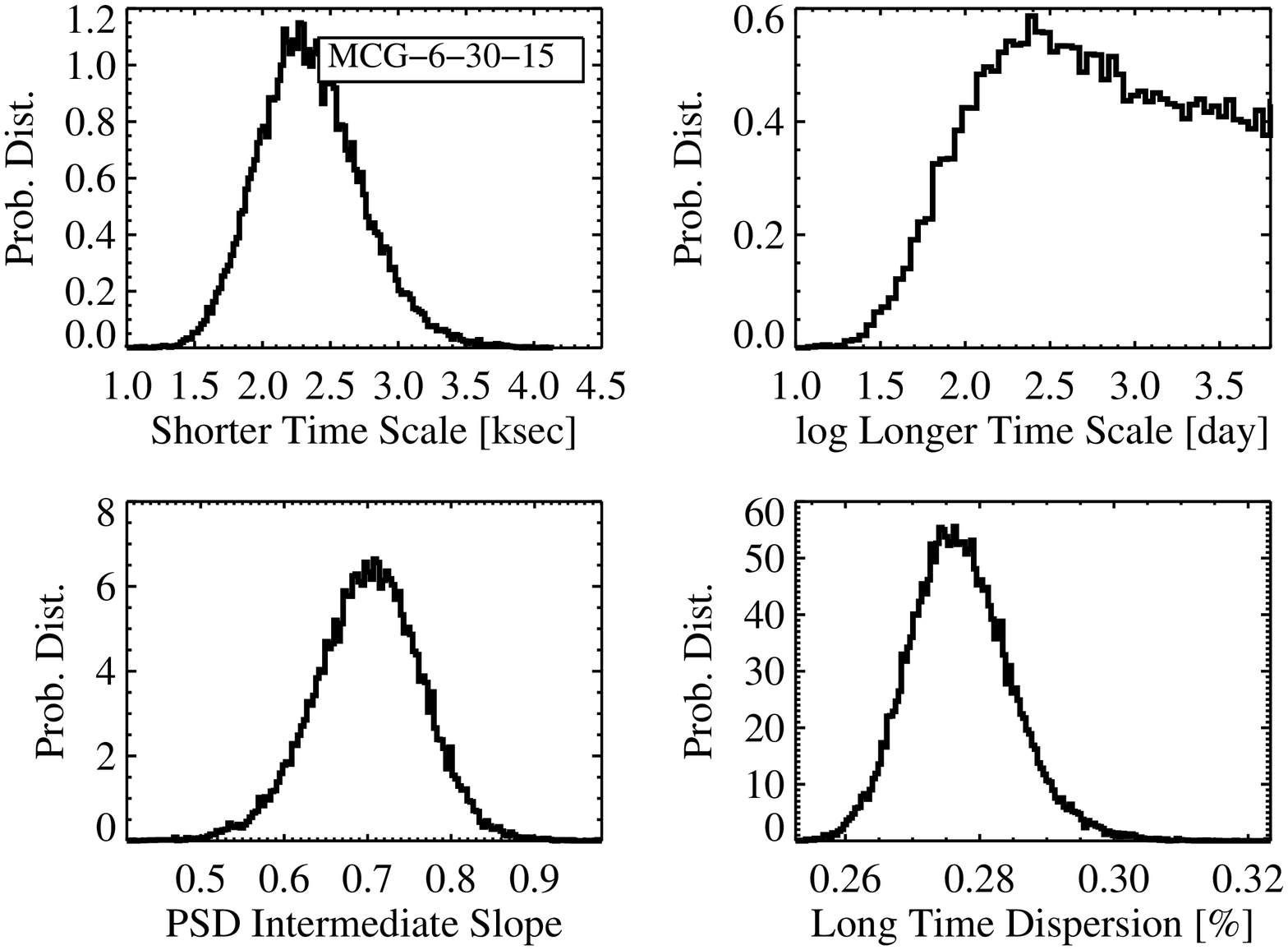}
    \caption{Posterior probability distribution for the short and long
      characteristic time scales, PSD slope between the break
      frequencies, and dispersion of the X-ray lightcurve for the AGN
      Akn 564 (left) and MCG-6-30-15 (right). For MCG-6-30-15 we are
      only able to place a lower limit on the longer characteristic
      time scale, or equivalently, a upper limit on the low-frequency break in the
      PSD. \label{f-agnpost}}
  \end{center}
\end{figure}

  In Figure \ref{f-agnpost} we show the posterior probability
  distributions of the characteristic time scales, logarithmic 
  slope of the power spectrum between the break frequencies, and the
  total dispersion in the fractional variations, for two
  sources. The first source, MCG-6-30-15, represents some of the best time
  coverage that we have for our sample. The second source, ARK 564, is
  the only AGN with evidence for two break frequencies in its power
  spectrum, and is thought to be analogous to galactic black holes in
  the `very high state' \citep{mchardy07}; all other AGN in our sample
  are thought to be analogous to galactic black holes in the `high/soft'
  state. We are able to confirm the existence of the two breaks in the
  power spectrum for ARK 564, corresponding to characteristic
  timescales of $\sim 300$ sec and $\sim 2$ days. However, for
  MCG-6-30-15, as with all other remaining AGN in our sample, there is
  still considerable probability on values of the longer time scale
  that are longer than the span of the lightcurve. As a result, we can
  only place a lower limit on the longer characteristic time scale, as
  we are only able to state with confidence that $\tau_L$ is not less
  than some value. For MCG-6-30-15, we find a lower limit of $\tau_L
  \gtrsim 37$ days.

\begin{figure}
  \begin{center}
    \includegraphics[scale=0.33,angle=0]{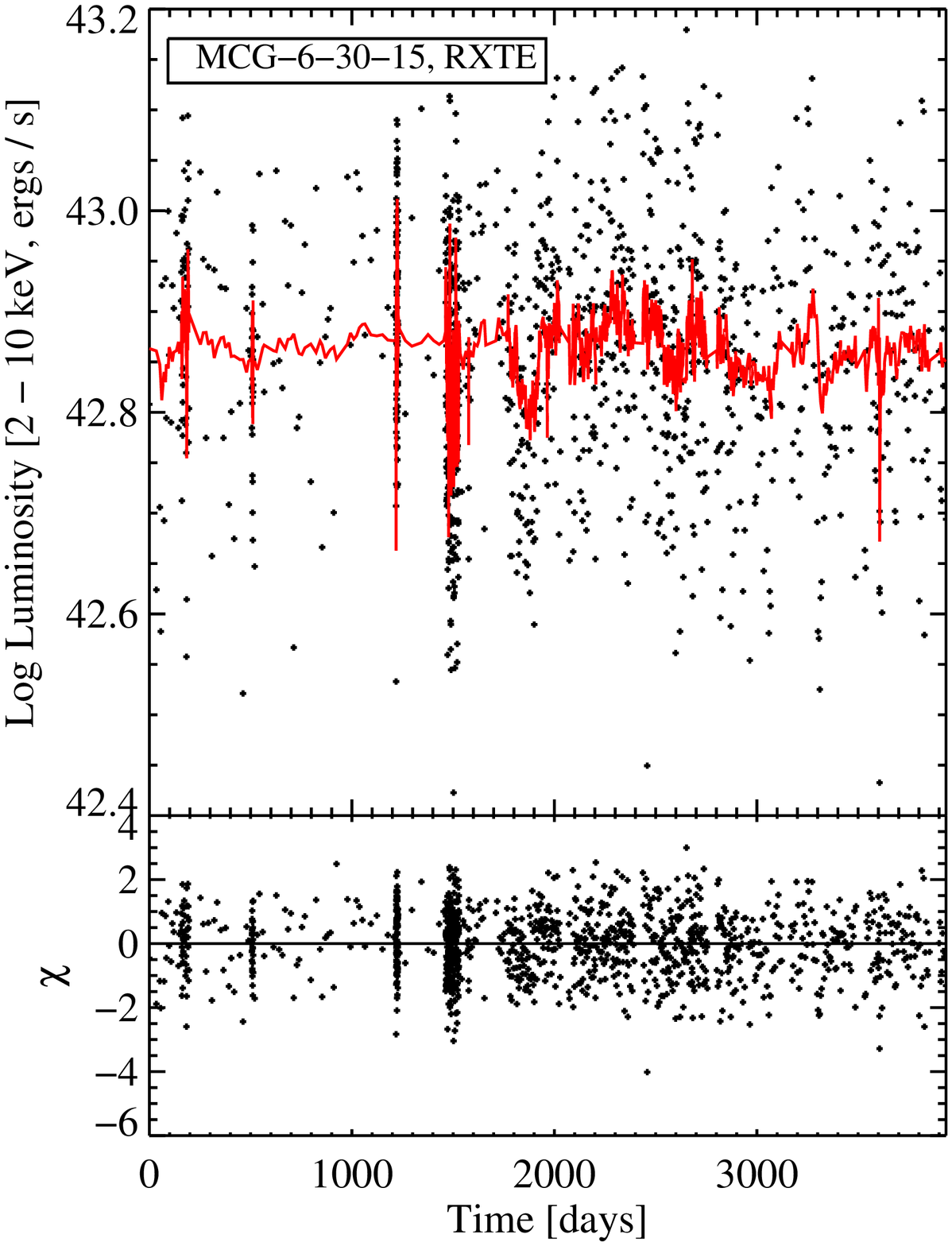}
    \includegraphics[scale=0.33,angle=0]{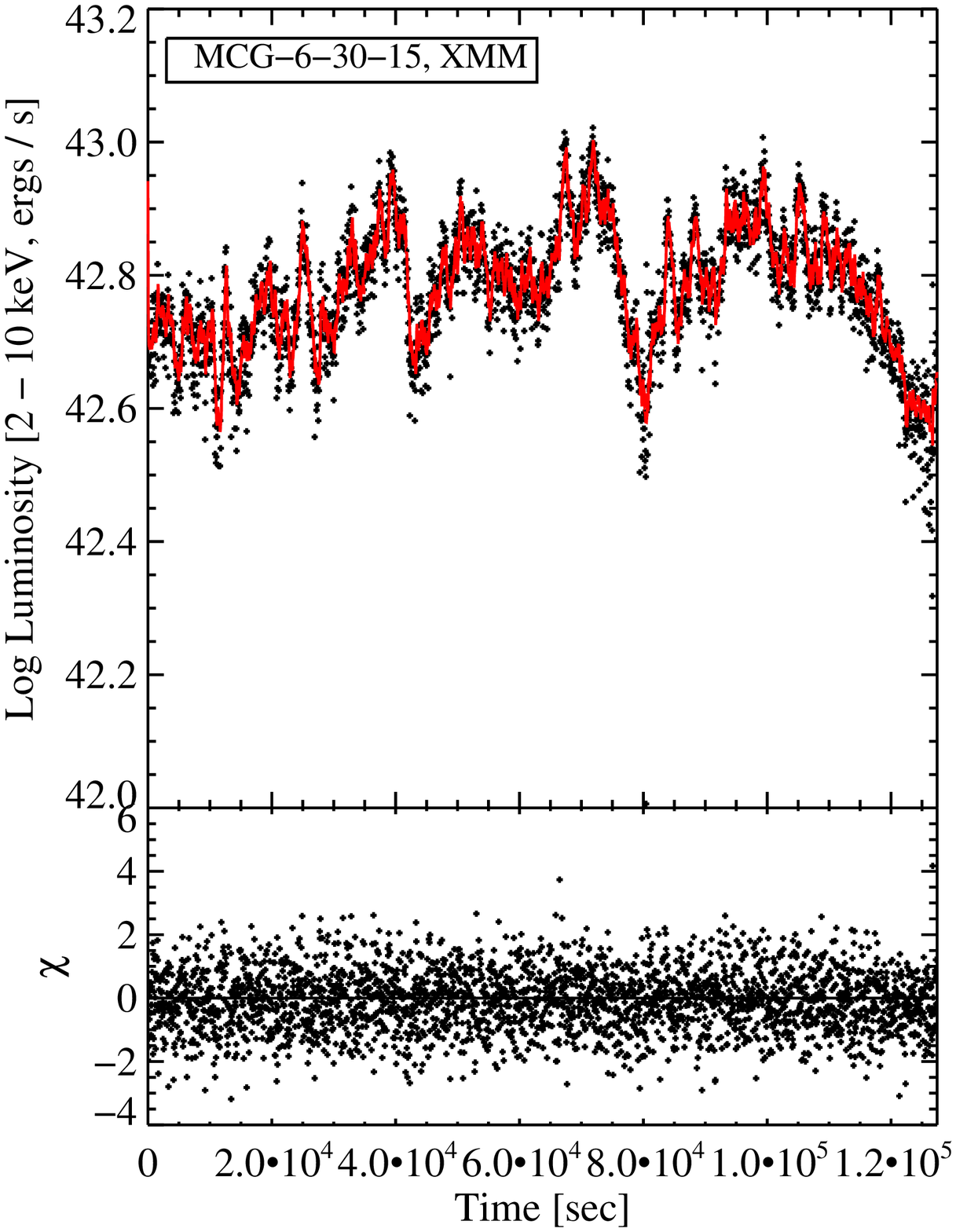}
    \caption{The observed RXTE (left) and XMM (right) lightcurves for MCG-6-30-15,
      compared with the running average of the best-fit mixed OU
      process model (red line) with power-law weights
      (Eq. [\ref{eq-plaw_weights}]). The standardized residuals, given
      by Equation (\ref{eq-sresids}) are also shown in the lower
      panels. The mixed OU process is able to provide a good fit to
      the X-ray lightcurve for MCG-6-30-15, as evidenced by the
      apparent white noise character of the residuals.
      \label{f-postcheck1}}
  \end{center}
\end{figure}

\begin{figure}
  \begin{center}
    \scalebox{1.0}{\rotatebox{0}{\plotone{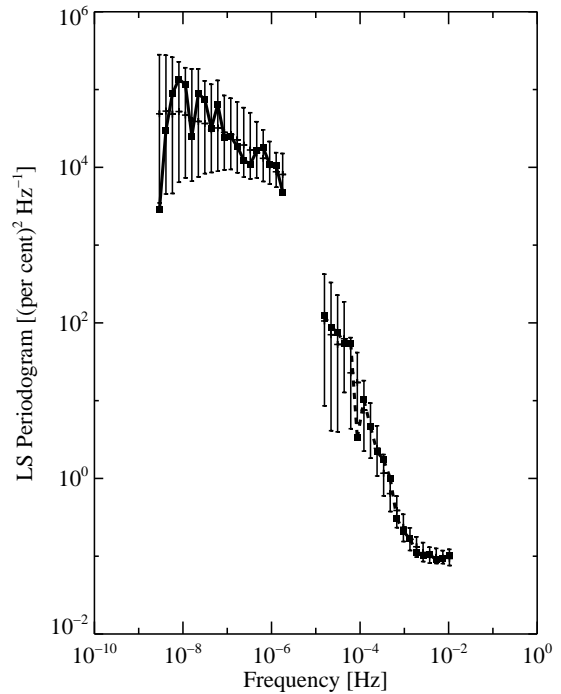}}}
    \caption{The Periodogram for the X-ray lightcurve of MCG-6-30-15
      (connected squares) compared with the distribution of
      periodograms expected under the mixed OU process model with
      power-law weights (points with error bars). The error bars
      encompass $90\%$ of the probability and represent the uncertainty
      in the expected observed periodogram due to uncertainty in the
      mixed OU process parameters, and due to variations in the
      sampled lightcurve due to its stochastic nature. The observed
      periodogram for 
      MCG-6-30-15 is consistent with that expected for the Mixed OU
      process model, confirming that this model can accurately model
      the variability of the X-ray fluctuations of this source across
      a large range in time scales.\label{f-postcheck2}}
  \end{center}
\end{figure}

  We assessed the goodness of fit using the checks described in
\S~\ref{s-goodfit}. In general, the mixed OU process provided a good
fit to the data, although the residuals exhibited some small
deviations from the assumed Gaussian distribution. We do not consider
this a concern, as it does not introduce a significant bias in the
estimated parameters. As an example, in Figure \ref{f-postcheck1} we
compare the lightcurve of MCG-6-30-15 with the running average for the
best-fitting mixed OU process. There do not appear to be any
systematic trends in the residuals accross the lightcurve. In
addition, in Figure \ref{f-postcheck2} we compare the periodogram of
the lightcurve for MCG-6-30-15 with that expected from the
best-fitting mixed OU process, after incorporating the uncertainty in
the model parameters. We used the Lomb-Scargle periodogram for the
RXTE data because of its irregular time sampling, while we used the
standard discrete Fourier transform for the XMM data. The mixed OU
process, with weights given by the bending power-law approximation
(Eq. [\ref{eq-plaw_weights}]), provides a good description of the
X-ray lightcurve of MCG-6-30-15, being able to model the variability
amplitude of luminosity fluctuations across a range of time scales.

\begin{figure}
  \begin{center}
    \includegraphics[scale=0.33,angle=90]{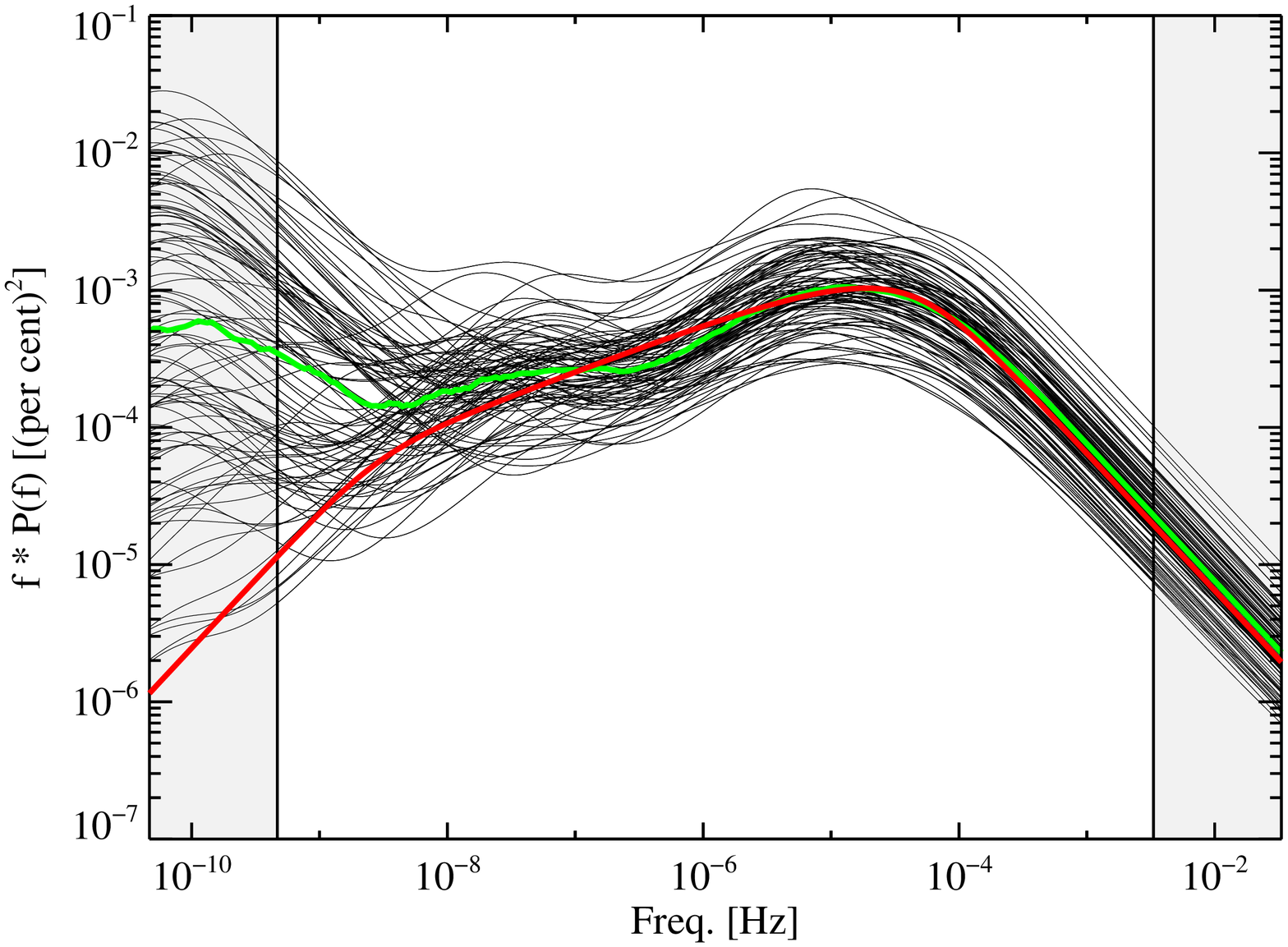}
    \includegraphics[scale=0.33,angle=90]{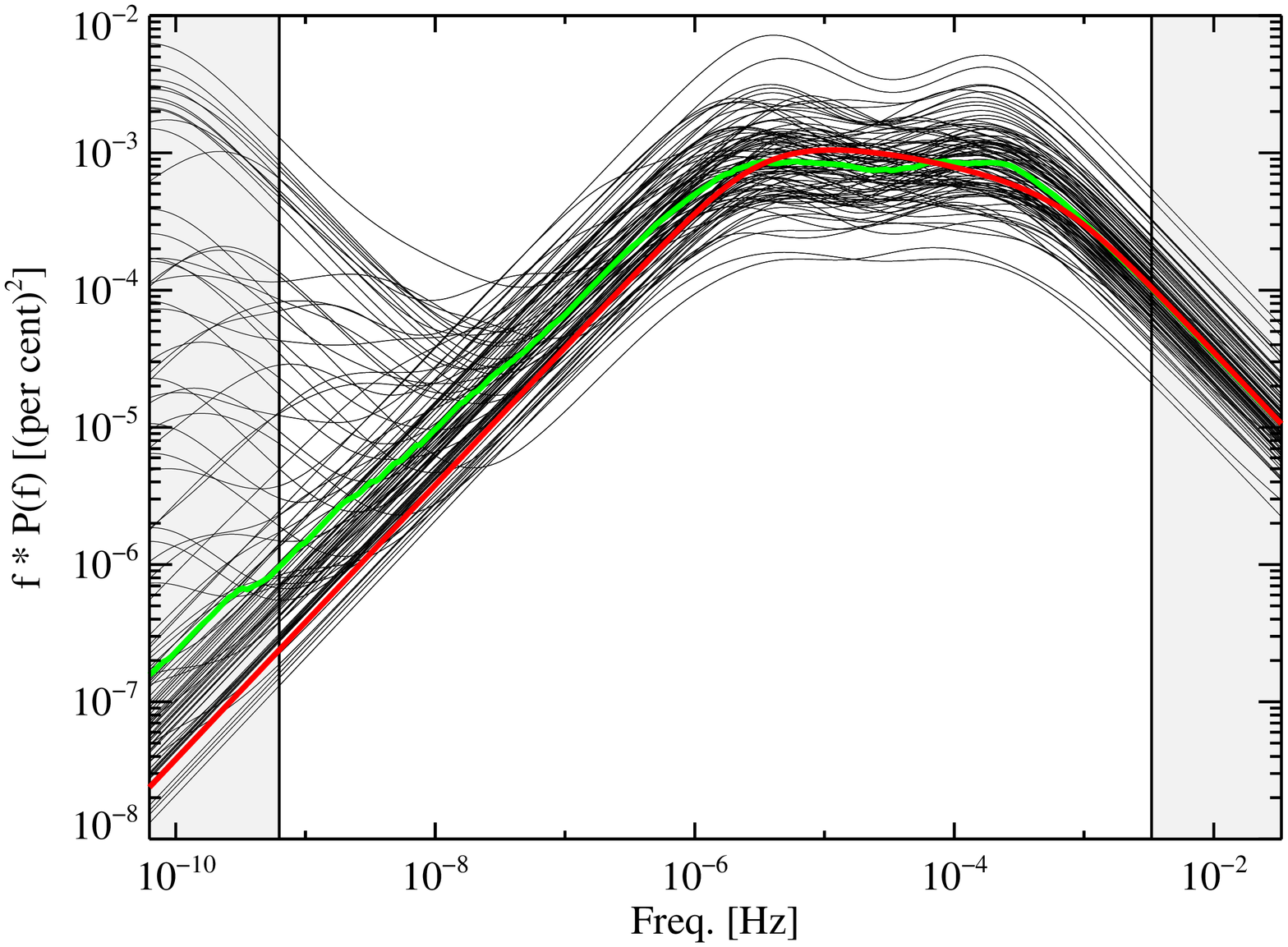}
    \caption{PSDs for MCG-6-30-15 (left) and Akn 564 (right) assuming
      the flexible weighting of the mixed OU process model. Here we
      have plotted $fP(f)$ for better comparison with other work. The
      thin black lines denote 100 random realizations of the PSD from
      its probability distribution, given the observed lightcurve, and
      the thick green lines denotes the median of the random
      realizations. The median value may be considered a `best-fit'
      value of the PSD, and the spread of the black lines give a
      graphical representation of the uncertainty in the PSD; that
      probability that the PSD has a certain value may be estimated by
      counting the fraction of black lines that intersect that
      value. The thick red line denotes the best-fit PSD assuming the
      power-law weighting, and the shaded regions denote the areas of
      frequency space not probed by the observed lightcurve. While the
      bending power-law provides a reasonable approximation to the
      PSDs, both sources display additional `wiggles' in the PSD that
      are not fit by the bending power-law model, similar to what is observed for GBHs.
      \label{f-xray_flexible}}
  \end{center}
\end{figure}

  As an additional check, we also used a flexible form for the mixed
OU process, treating the values of the 32 weights as free parameters, but the break
frequencies being fixed on a logarithmic grid in $\omega_j$. This
model provides more flexible modeling of the PSD, and we fit both the
Akn 564 and MCG-6-30-15 X-ray lightcurves with it. The results are
shown in Figure \ref{f-xray_flexible}, where we compare the `flexible'
weighting with the power-law weighting. Note that here we have
multiplied the PSDs by frequency, for easier comparison with
previous work. We can still get good constraints on the PSD with the
flexible weighting, although the uncertainties are now larger. For
MCG-6-30-15, there does not appear to be a significant difference in
the estimated PSD assuming the flexible weighting compared
to the power-law weighting, although there may be some additional
`wiggles' in the PSD which the power-law weighting cannot
fit. Similarly, the estimated PSD for Akn 564 assuming the flexible
weighting displays some wiggles, suggesting that the simple bending
power-law model is not able to capture all of the features of the
PSD. Indeed, \citet{mchardy07} find that a sum of Lorentzians provides
a better fit to the PSD of Akn 564. Moreover, 'wiggles' and other
deviations from a power-law PSD are seen in the X-ray PSDs of GBHs,
and thus we might also expect them for AGN.

  \subsubsection{Comparison with Black Hole Mass}

  \label{s-mbh_compare}
  
\begin{figure}
  \begin{center}
    \scalebox{0.8}{\rotatebox{90}{\plotone{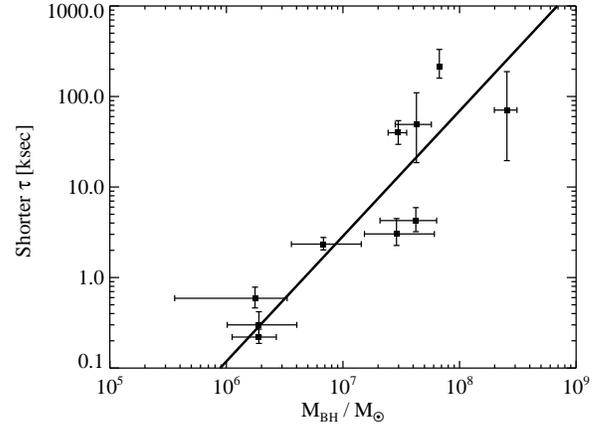}}}
    \caption{Dependence of characteristic time scale corresponding to
      the X-ray high frequency break on black hole mass. The error
      bars denote $68\%$ confidence regions, and the solid line shows
      the best-fit linear relationship
      (Eq.[\ref{eq-mbh_tau}]). Similar to previous work, we find a
      clear dependence of $\tau_{H}$ on $M_{BH}$. \label{f-mbh_tau}} 
  \end{center}
\end{figure}

  \citet{sob09b} compiled black hole masses for the AGN, most of which
  are based on reverberation mapping \citep{peter04}, and we use the
  values tabulated in \citet{sob09b}. The only exception is Mrk 766, which
  has a mass based on reverberation mapping calculated by \citet{bentz09}.
  In Figure \ref{f-mbh_tau} we plot the characteristic time scale
  corresponding to the high frequency break, $\tau_H$, as a function
  of black hole mass. We find a strong correlation between the two
  quantities, although this is not surprising as several authors have
  previously found a correlation based on direct PSD fitting of these
  same sources \citep[e.g.,][]{uttley05a,mchardy06}. However, using
  our statistical technique we obtain estimates for the break time
  scales for the two highest mass objects, Fairall 9 and NGC 5548,
  which previous work did not include as only lower limits on the
  break time scales were available. This correlation has typically
  been interpreted as being driven by a correlation between black hole
  mass and some relevant physical time scale in the X-ray emitting
  region, e.g., the viscous or thermal time scale, which corresponds
  to the suppression of fluctuations on time scales short compared to
  this characteristic time scale. However, 
  based on the discussion in \S~\ref{s-oudiff}, another, possibly
  related, interpretation is that the long time scale break
  corresponds to the characteristic time scale of the physical process
  driving the fluctuations, which increases with black hole mass, while
  the correlation between the short time scale break and black hole
  mass is caused by an increase in the characteristic spatial scale of 
  the noise process driving the fluctuations with increasing black hole mass. 

  We fit a linear relationship between $\log \tau_{H}$ and $\log M_{BH}$ using the
  Bayesian linear regression method of \citet{kelly07}, which accounts
  for the measurement errors in both $\log \tau_H$ and $\log
  M_{BH}$. It is especially important in this case to use a Bayesian
  method which incorporates 
  the measurement errors because we only have 10 data points, and thus
  Bayesian methods are needed in order to accurately quantify the
  uncertainties. We find that on average
  \begin{equation}
    \tau_H = 3.00^{+4.57}_{-1.90} \left(\frac{M_{BH}}{10^7
        M_{\odot}}\right)^{1.39 \pm 0.64} ({\rm ksec})
    \label{eq-mbh_tau}
  \end{equation}
  where the errors are quoted at $90\%$ confidence. We estimate the
  intrinsic scatter in $\tau_H$ at fixed $M_{BH}$ to 
  be $\sim 0.6$ dex. There is still potentially large intrinsic
  scatter in $\tau_H$ at fixed $M_{BH}$, which is likely due to
  scatter in the accretion rates at fixed $M_{BH}$
  \citep[e.g.,][]{mchardy06}; however, we stress that the estimate of
  the dispersion in the intrinsic scatter is highly uncertain, and
  more data is needed.

\begin{figure}
  \begin{center}
    \scalebox{0.8}{\rotatebox{90}{\plotone{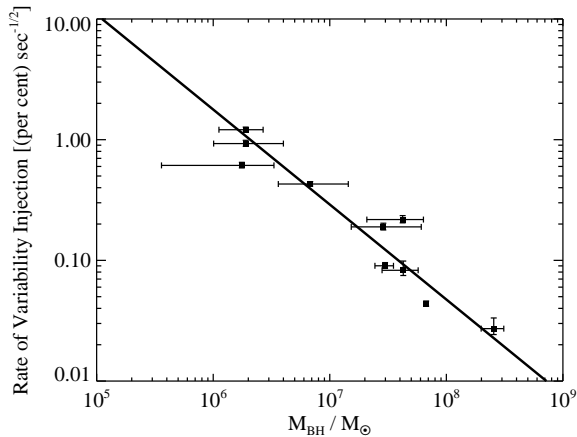}}}
    \caption{Dependence of $\varsigma$ on black hole mass; note that
      $\varsigma$ parameterizes the amplitude of the X-ray PSD on time scales
      short compared to the high frequency break, and is proportional
      to the amplitude of the driving noise field for the mixed OU
      process model. The error
      bars denote $68\%$ confidence regions, and the solid line shows
      the best-fit linear relationship
      (Eq.[\ref{eq-mbh_tau}]). There is a clear, very tight
      relationship between $\varsigma$ and $M_{BH}$. \label{f-mbh_sig}} 
  \end{center}
\end{figure}

  We also searched for correlations of $M_{BH}$ with the PSD slope
  below the high frequency break $\alpha$,
  amplitude of driving noise $\varsigma$, and the total variability amplitude based
  on the mixed OU process model. We did not find any significant
  evidence for trends involving $\alpha$ or the total variability amplitude;
  however, we did find a highly significant anti-correlation between
  $\varsigma$ and $M_{BH}$, as shown in Figure \ref{f-mbh_sig}. The trend
  is surprisingly tight, and implies that the fractional X-ray variations on time scales
  short compared to the high frequency break are weaker for AGN with
  more massive black holes (see Eq.[\ref{eq-highf_var}]), or rather that the rate at which variability power is injected into the lightcurve decreases with increasing $M_{BH}$. Within the context of the stochastic
  linear diffusion model, this implies that the fractional amplitude of the
  driving noise field, which may be associated with MHD turbulence,
  decreases with $M_{BH}$. The Bayesian linear regression finds
  \begin{equation}
    \varsigma = 0.29^{+0.14}_{-0.08} \left(\frac{M_{BH}}{10^7
        M_{\odot}}\right)^{-0.79 \pm 0.22} ({\rm per\ cent\ sec}^{-1/2})
    \label{eq-mbh_sig}
  \end{equation}
  where the error are quoted at $90\%$ confidence. We estimate the
  intrinsic scatter in $\varsigma$ at fixed $M_{BH}$ to 
  be $\sim 0.2$ dex.

 This result is in agreement with previous work which has found an anti-correlation between the excess variance in the X-ray lightcurve and $M_{BH}$ \citep[e.g.,][]{lu01,pap04,nik04,oneill05,zhou10}. A similar anti-correlation between the rate at which variability power is injected into the lightcurve and black hole mass has also been found for AGN optical variations (KBS09). \citet{czerny01} find an anti-correlation between $M_{BH}$ and the frequency at which the PSD reaches a reference value, which is consistent with an anti-correlation between $M_{BH}$ and the amplitude of the high-frequency PSD. However, our result differs from most previous work in that we do not compute the variance over some range of time scales, directly from the observed lightcurve, but rather we fit a parameter $\varsigma$, which is the rate at which variability power is injected into the lightcurve. As shown in Equation (\ref{eq-highf_var}), the parameter $\varsigma^2$ is proportional to the short time scale variance, but is not strictly equal to it, as it does not depend on the time scales probed by the observed lightcurve. Indeed, under our adopted normalization for the mixed OU process weights (Eq.[\ref{eq-unitnorm}]), $\varsigma^2$ essentially gives the normalization of the PSD on time scales short compared to the high frequency break. While the value of $\varsigma$ can be precisely determined in an unbiased manner using our statistical method, this is not the case for the observed variance in a lightcurve over some range of time scales. The observed variance in a lightcurve over some range of time scales is often a poor estimate of the actual time-averaged variance \citep{vaughan03,oneill05}. The situation worsens if one is estimating the variance in the lightcurve using different time scales for different sources.

While we have shown that the ampitude of the high frequency variance
  is anti-correlated with $M_{BH}$, physically interpreting this
  correlation is difficult. Within the context of our mixed OU process
  model the anti-correlation implies that the amplitude of the driving
  noise field, and thus the rate at which variability power is
  injected into the lightcurve, decreases with increasing
  $M_{BH}$. However, this is also conditional on our adopted
  normalization for the mixing weights (Eq.[42]), and,
  as we stated earlier, the absolute values of the mixing weights are
  degenerate with $\varsigma^2$. Under our normalization, an
  anti-correlation between $\varsigma^2$ and $M_{BH}$ would be
  expected if the value of the PSD in between the break frequencies
  is independent of $M_{BH}$, assuming the PSD slope is equal to $-1$, as
  $\omega_H$ is also anti-correlated 
  with $M_{BH}$. Thus, the observed anti-correlation between
  $\varsigma^2$ and $M_{BH}$ could be a manifestation of the fact that
  the variability amplitude per frequency interval is constant between the break
  frequencies, but $\omega_H$ decreases with increasing
  $M_{BH}$. However, if it were true that the $\varsigma$--$M_{BH}$
  anti-correlation is merely an artifact of the $\tau_H$--$M_{BH}$
  correlation, then we would expect the $\varsigma$--$M_{BH}$
  relationship to have a slope of $-0.5$. This is because the
  parameter $\varsigma^2$ is proportional to the value of the PSD at
  fixed frequency, provided that $\omega \gg \omega_H$. If the value
  of $\omega P(\omega)$ is constant at the high frequency break, and
  if the break frequency scales as $\omega_H \propto M^{-1}_{BH}$, then
  we would expect that $\varsigma \propto M_{BH}^{-1/2}$. However,
  the dependence of $\varsigma$ on $M_{BH}$ that we find in Equation
  (51, $\varsigma \propto M_{BH}^{-0.79}$) is steeper than that expected under this assumption of a scale-invariant PSD
  shape. This implies that the $\varsigma$--$M_{BH}$
  anti-correlation is not simply an artifact of the $\tau_H$--$M_{BH}$
  correlation, but is real.

\begin{figure}
  \begin{center}
    \scalebox{0.8}{\rotatebox{90}{\plotone{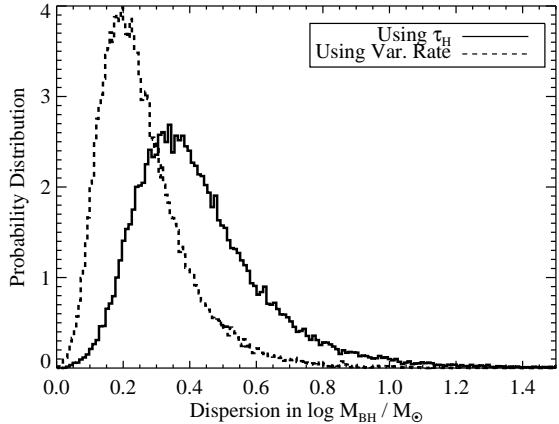}}}
    \caption{Probability distribution for the scatter in mass at a
      given $\tau_{H}$ (solid curve) or $\varsigma$ (dashed curve),
      given the 10 AGN in our data set. For comparison, mass estimates obtained
      from scaling relationships involving the broad emission lines
      and the host galaxy properties are $\sim 0.4$ and $\sim 0.3$
      dex, respectively, implying that $\varsigma$ may give the most
      precise single-observation mass estimates. \label{f-masserr_prob}}
  \end{center}
\end{figure}

  The correlations between the high frequency break or $\varsigma$ and the black hole
  mass raise the possibility of using the X-ray lightcurve as an
  additional method for estimating black hole mass for AGN. It is thus
  of interest to assess the accuracy of the estimates derived from
  $\tau_H$ or $\varsigma$. To do this, we also performed a linear regression of
  $\log M_{BH}$ on $\log \tau_H$ and $\log \varsigma$, respectively. Note that we cannot simply invert
  Equations (\ref{eq-mbh_tau}) and (\ref{eq-mbh_sig}), but need to perform the regression after
  switching the variables. Based on our results, mass estimates may be
  obtained as
  \begin{eqnarray}
    \log (M_{BH} / M_{\odot}) & = & (6.75 \pm 0.20) + (0.60 \pm 0.14) \log
    (\tau_{H} / [1\ {\rm ksec}]) \label{eq-mass_estimate1} \\
    \log (M_{BH} / M_{\odot}) & = & (6.36 \pm 0.19) - (1.20 \pm 0.18) \log
    (\varsigma / [({\rm per\ cent})\ {\rm sec}^{-1/2}]) \label{eq-mass_estimate2}
  \end{eqnarray}
  Here, unlike before, we have quoted the errors at $68\%$ ($1\sigma$)
  confidence. The dispersion in $\log M_{BH}$ at fixed $\tau_H$ is $\sim 0.4$ dex,
  and the dispersion in $\log M_{BH}$ at fixed $\varsigma$ is $\sim 0.2$
  dex, in agreement with the estimate of \citet{zhou10} when using the
  total high frequency variability amplitude. 
  The probability distribution of the dispersion in mass at fixed
  $\tau_{H}$ and $\varsigma$ is shown in Figure \ref{f-masserr_prob};
  this quantity gives the precision in using these quantities as
  estimates of black hole mass. These results imply that mass
  estimates obtained from the high frequency break 
  have similar accuracy to those obtained from the broad emission
  lines \citep[$\sim 0.4$ dex,][]{vest06}, but are slightly less
  accurate than those obtained from the
  $M_{BH}$--$\sigma$ relationship \citep[$\sim 0.3$
  dex,][]{novak06,gult09a}. The mass estimates obtained from $\varsigma$, on the
  other hand, appear to be the most accurate, on average, of all the
  single-observation techniques, at least for AGN, and have the advantage
  that the parameter $\varsigma$ can be determined from a single
  X-ray lightcurve with significantly higher precision than can the break
  frequency.

  \subsection{AGN Optical Lightcurves}

  \label{s-optagn}

  Recent work has suggested that the optical lightcurves of AGN are well
  described by a single OU process
  \citep{kbs09,koz10,macleod10a}. The power spectrum for a single OU
  process becomes flat below the break frequency. It is
  worth investigating whether there is evidence for an intermediate
  region below the break frequency where the power spectrum falls off
  as $1 / f^{\alpha}, (-2 < \alpha < 0)$, similar to the X-ray
  lightcurves. We can do this by testing whether the optical
  lightcurves are better described by a mixed OU process, instead of a
  single OU process. We used a mixed OU process with weights given
  by Equation (\ref{eq-plaw_weights}) to fit the high
  quality optical
  lightcurves of 63 AGN from the \emph{MACHO} \citep{geha03} and AGN Watch samples of
  \citet{kbs09}. The 55 \emph{MACHO} $R$- and $V$-band lightcurves
  typically have seasonal time sampling of $\sim 2$--$10$ days over
  $\sim 7.5$ years. The 8 Seyfert Galaxies from the AGN Watch database
  have optical lightcurves with varied time sampling and lengths, and
  were monitored for reverberation mapping. Further details of these
  samples are described in \citet{kbs09} and references therein.

  We searched for objects for which the mixed OU process
  provided a better fit by finding those objects satisfying the
  union of the following two criteria:
  \begin{itemize}
  \item
    Most of the posterior probability for the ratio of break frequencies was
    found at values $\omega_H / \omega_L > 1$.
  \item
    Most of the posterior probability for the slope of the PSD between
    the two break frequencies was found at values $-2 < \alpha < 0$.
  \end{itemize}
  Note that these criteria essentially amount to finding the set of PSDs which do
  not reduce to the Lorentzian shape of the single OU process. For
  most sources, we did not find significant evidence that the 
  mixed OU process provided a better fit to AGN optical variability
  than did a single OU process. Moreover, the characteristic time
  scales became significantly more uncertain under the mixed OU
  process model, most likely because we allowed the slope above the high
  frequency break to vary. However, we did find that for seven of the
  \emph{MACHO} sources the mixed OU process provided a better fit
  to both the $R$- and $V$-band lightcurves. These sources had optical
  PSDs consistent with a single unbroken power-law with slope $-2 <
  \alpha < 0, P(f) \propto f^{\alpha}$, and thus no break frequencies
  were detected. There were a few
  additional objects for which either the $R$- or $V$-band PSD
  deviated from the Lorentzian shape of the OU process, but 
  not both. 

  The seven objects for which both the $R$- and $V$-band
  lightcurves were better described by the mixed OU process are
  summarized in Table \ref{t-optical}. We could not find anything
  unusual about these objects other than their timing properties. We
  also fit a flexible form of the mixed OU process to some of the better
  sampled optical lightcurves, treating the weights as free parameters
  on a fixed logarithmic grid for $\omega_j$. We could not find any
  significant evidence that the optical PSDs diverged from either a single
  power-law or Lorentzian PSD.

\begin{figure}
  \begin{center}
    \scalebox{0.8}{\rotatebox{90}{\plotone{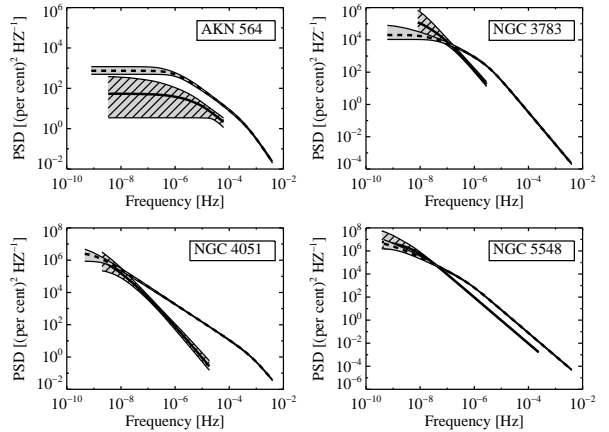}}}
    \caption{Comparison of the optical (solid line, diagonal line
      fill) and X-ray (dashed line, solid fill) PSDs for the four
      objects in our sample for which we have both good X-ray and
      optical lightcurves. The lines denote the best-fit PSDs and the
      shaded regions denote the $68\%$ confidence regions for the
      PSDs, assuming the mixed OU process model with power-law
      weighting. In general, the optical PSDs fall off faster toward
      higher frequencies, implying that the short time scale
      variations are more suppressed in the optical
      lightcurves. However, the optical lightcurves appear to have
      more variability on longer time scale than the X-ray
      lightcurves, implying that the optical variations are not solely
      due to reprocessing of the X-ray emission. The only exception is
      Akn 564, which has a high accretion rate and is thought to be
      analogous to galactic black holes in the very high
      state.\label{f-optxray_psd}} 
  \end{center}
\end{figure}

  We have both optical and X-ray lightcurves for Akn 564, NGC 3783,
  NGC 4051, and NGC 5548, and it is worth comparing the two estimated PSDs for
  each source. In Figure \ref{f-optxray_psd} we compare the estimated
  optical and X-ray PSDs for each source assuming the mixed OU model
  with power law weighting. For all sources we subtracted off the host
  galaxy flux, where we used the values reported by \citet{bentz09}
  for NGC 3783, NGC 4051, and NGC 5548, and the value reported by
  \citet{shem01} for ARK 564. In general, the optical fluctuations
  are less variable then the X-ray fluctuations, except possibly on time
  scales $\gtrsim 100$ days. Furthermore, the optical PSDs fall off
  steeper toward high frequencies than the X-ray PSDs, and thus the
  short time scale optical fluctuations are more strongly suppressed
  relative to the X-ray fluctuations. These results are consistent with
  previous comparisons of optical and X-ray PSDs
  \citep[e.g.,][]{czerny99,czerny03,uttley03,arev08,arev09,breedt09,breedt10}.

  One possible interpretation of these results is that the different
PSDs represent how the different media in the two regions respond to a
common driving process that is turbulent in both space and time, with
the region emitting the optical photons more strongly suppressing the
short-time scale noise fluctuations. This is consistent with the fact
that $\varsigma$ is anti-correlated with $M_{BH}$ for both the X-ray
and optical lightcurves. For example, within the propagation class of
models, this implies that while the source of the accretion rate
perturbations may be the same, the optical emitting region more
strongly suppresses the short time scale accretion rate perturbations,
but when these perturbations reach the X-ray emitting region (i.e.,
the corona), which may also be subject to a driving noise field, the
short time scale fluctuations are not as strongly filtered as this
medium is not as `stiff'. However, if the optical PSDs continue to
rise above the X-ray PSDs at longer time scales, as it appears they
do, then the total amplitude of variability in the optical
fluctuations surpasses that in the X-ray fluctuations. This is a
departure from the behavior seen in soft state GBH, where the disk
emission is significantly less variable than the corona, suggesting a
descrepancy between the spectral-timing properties of GBHs and AGN. In
addition, larger optical variability amplitude than X-ray variability
amplitude implies that either the source of the variations is
different for the two emission regions, or that the power in the
disturbances is somehow dissipated such that their variability
amplitude has decreased by the time they reach the X-ray emitting
region. However,
the results also imply that all of the optical variability cannot be
due to reprocessing of X-ray emission, as has been shown in previous
work \citep[e.g.,][]{czerny99,czerny03,uttley03}. The only possible
exception appears to be Akn 564, where the X-ray variability amplitude
is larger at all time time scales. Akn 564 is noteworthy as it is the
only AGN known to exhibit a second low-frequency break, and the
additional X-ray variability may be due to an increase in the strength
of the X-ray power-law component, such as is seen in GBH in the VHS.

  \section{SUMMARY}

  \label{s-summary}

  In this work we have developed a new model for the lightcurves of
  accreting black holes. The model is based on a mixture of
  Ornstein-Uhlenbeck stochastic processes and also happens to be the
  solution to the linear diffusion equation perturbed by a spatially
  correlated noise field. In essence, our model for quasar lightcurves
  may be thought of as a type of basis expansion, where the basis for
  the lightcurve is a set of independent stochastic processes. This allows
  flexible modeling of the PSD so long as $-2 \leq d\log P / d\log f
  \leq 0$ for all frequencies of interest, and we show that the mixed
  OU process provides a good approximation to the X-ray lightcurves of
  GBHs, and the optical and X-ray lightcurves of AGN. Our main results
  are 
  \begin{itemize}
  \item
    The mixed OU process is the solution to the linear diffusion
    equation perturbed by a spatially correlated noise field, and
    thus the mixed OU process describes the evolution of viscous,
    thermal, or radiative perturbutions due to a driving noise, to
    the extant that the viscous, thermal, or radiative response of
    the accretion flow follows a linear diffusion equation. Within
    this interpretation, the low frequency break in the PSD
    corresponds to the diffusion time scale in the outer region of
    the accretion flow, and the shape of the PSD above the low frequency
    break depends on the viscosity of the flow (or, more generally,
    the diffusion coefficient). If the noise field is spatially
    correlated, then an additional high frequency break in the PSD
    may exist at the frequency corresponding to the time it takes a perturbation
    traveling at the viscous speed to cross the characteristic
    spatial scale of the noise field. 
  \item
    We derive the likelihood function of a lightcurve for the
    mixed OU process, given by Equations
    (\ref{eq-lik})--(\ref{eq-deltai}). This allows one to estimate
    the parameters of the mixed OU process for an observed
    lightcurve, and equivalently the PSD, in a manner which is
    computationally efficient, unbiased by red noise leak and
    aliasing, and fully accounts for irregular sampling and
    measurement errors. If the parameters for a bending power-law
    are desired, then we also derive a form for the mixing weights
    which approximates a bending power-law, given by Equation
    (\ref{eq-plaw_weights}). 
  \item
    We show that the mixture of Lorentzians PSDs implied by the mixed
    OU process stochastic model is a good fit to the PSD of an
    observation of Cygnus X-1 in the hard state.
  \item
    We applied our mixed OU process model to the X-ray lightcurves
    of 10 well-studied local AGN and show that it provides a good
    description of their lightcurves, and we recover many of the PSD
    features which have been obtained previously from direct fitting
    of the observed PSDs. We use the estimated break frequencies to
    recover the correlation between time scale of the high frequency
    break and black hole mass seen previously with this sample; this
    includes new estimates of the high frequency time scales for
    Fairall 9 and NGC 5548, for which previously only upper limits
    were obtained. 
  \item
    We find a significant anti-correlation between black hole mass and
    the amplitude of the driving noise fluctuations, 
    $\varsigma$, which also parameterizes the amplitude of the high frequency
    PSD. The form of this correlation is similar to what has been
    observed in optical lightcurves (KBS09), suggesting that the
    origin of the optical and X-ray variability for AGN is at least
    partially shared. This correlation is even tighter than that 
    between $M_{BH}$ and the break frequency. Mass estimates obtained
    using $\varsigma$ appear to have an uncertainty of $\sim 0.2$ dex,
    potentially making it the best single-observation mass estimator for AGN.
  \item
    We did not find significant evidence that most optical
    lightcurves of AGN deviated significantly from a single OU
    process, i.e., a Lorentzian PSD. However, we did find
    evidence that the optical lightcurves of 7 AGN in a sample of
    55 AGN from the \emph{MACHO} survey exhibited PSDs which were
    flatter than the $1 / f^2$ shape seen in the majority of the
    sample, but the origin of this difference is unclear.  These sources may exhibit flatter PSDs because we are seeing them on the $1 / f^\alpha$ part of the PSD, i.e., the break time scales are shorter and longer than the minimum and maximum time scales probed by the lightcurves.
  \item
    We compare the optical and X-ray PSDs for Akn 564, NGC
    3783, NGC 4051, and NGC 5548 and find that for the three
    NGC sources the short time scale fluctuations are more
    suppressed in the optical, but the long time scale
    fluctuations appear to be stronger in the optical. This
    implies that the total variability amplitude in the optical
    fluctuations is larger than that in the X-ray fluctuations,
    and thus the optical fluctuations cannot be caused solely
    by reprocessing of X-ray emission. In addition, if both the
    optical and X-ray fluctations are caused by inwardly
    propagating fluctuations in, say, the accretion rate, then
    some of the power in the fluctuations must be dissipated by
    the time they reach the X-ray emitting region. Akn 564, the
    only AGN believed to be in the very high accretion rate state, does
    not exhibit this behavior, and the optical fluctuations are
    always less variable then the X-ray fluctuations, at least
    over the range of time scales we probe. 
  \end{itemize}

  \acknowledgements

  We would like to thank Phil Uttley for helpful conversations, and Simon Vaughan, Christopher Kochanek, Chelsea MacLeod, and Eric Feigelson for looking over and commenting on an earlier version of this manuscript. We would also like to thank the anonymous referee for a careful reading and detailed comments which improved the quality of this paper. BK acknowledges support by NASA through Hubble Fellowship
grants \#HF-01220.01 and \#HF-51243.01 awarded by the Space Telescope
Science Institute, which is operated by the Association of
Universities for Research in Astronomy, Inc., for NASA, under contract
NAS 5-26555. This research is funded in part by NASA contract NAS8-39073.  Partial
support for this work was provided by the {\it Chandra\/} grant
GO8-9125A. This research has made use of data obtained from the High Energy Astrophysics Science Archive Research Center (HEASARC), provided by NASA's Goddard Space Flight Center. This research has also made use of the NASA/IPAC Extragalactic Database (NED) which is operated by the Jet Propulsion Laboratory, California Institute of Technology, under contract with the National Aeronautics and Space Administration.

  \appendix

  \section{L\'{e}vy Processes}

  L\'{e}vy processes form a general set of processes that include
  Brownian motion and compound Poisson processes, both relevant in
  astrophysics. Basically, a stochastic process is a L\'{e}vy process
  if it has stationary and independent increments. Brownian motion, or
  rather the Wiener process, is a random walk process with a power 
  spectrum that decays as $1 / \omega^2$. A white noise process is the
  derivative of Brownian motion, and is a stochastic process with mean
  zero and a flat power spectrum; conversely, Brownian motion is the
  integral of white noise. A Brownian motion for the process driving
  the luminosity fluctuations might 
  be considered a reasonable approximation if the input driving noise,
  $dW(t)$, has a decorrelation time
  scale short compared to the characteristic time scale of
  the luminosity fluctuations, therefore implying that $dW(t)$ resembles white
  noise. Alternatively, a compound Poisson process may be considered
  to describe the variability in luminosity caused by flares above an
  accretion disk. If the number of flares in a time interval follows a
  Poisson distribution, and the luminosities of the flares are
  randomly drawn from some other probability distribution, then the
  sum of the flare luminosities is a compound Poisson process. In
  general, most of the results from \S~\ref{s-smodel} are valid for any L\'{e}vy
  process with zero mean and unit variance. The assumption that the driving noise,
  $dW(t)$, has zero mean and unit variance is not very restrictive, as
  $\mu$ and $\varsigma$ can always be rescaled to make this so.

  \section{Continuous Mixtures of Ornstein-Uhlenbeck Processes}

  A continuous mixture of
  independent OU processes, denoted as $Y(t)$, may be obtained as
  \begin{equation}
    Y(t) = \mu + \int_{0}^{\infty} \int_{0}^{\infty} X(t,\omega_0,\varsigma) C(\omega_0,\varsigma)\
    d\omega_0\ d\varsigma
    \label{eq-mixou}
  \end{equation}
  Here, $C(\omega_0,\varsigma)$ is a weighting function which describes
  how much a given OU process contributes to the mixture, and the
  notation $X(t,\omega_0,\varsigma)$ defines an infinite set of
  independent OU processes indexed by the value of $\omega_0$ and
  $\varsigma$; we will refer to 
  $C(\omega_0,\varsigma)$ as the `mixing function'.  We assume that the
  individual OU processes have zero 
  mean, and that the mean of the mixed OU process is $\mu$.  Some
  previous work has focused on the special case where
  $C(\omega_0,\varsigma)$ is a probability distribution
  \citep[e.g.,][]{igloi99,eliazar09}. In general, the process $Y(t)$
  is not Markovian but stationary \citep{eliazar09}.

  We are allowed considerable flexibility in modeling the
  lightcurve $Y(t)$ via the mixing function, and as an example we focus on
  the special case where $C(\omega_0,\varsigma)$ is a step function over some
  range of $\omega_0$. Specifically,
  \begin{equation}
    C(\omega_0,\varsigma) = \left\{ \begin{array}{l}
        1, \omega_{L} \leq \omega_0 \leq \omega_{H} \\
        0, {\rm otherwise} \end{array} \right. .
    \label{eq-mixfunc}
  \end{equation}
  The autocovariance function and power spectrum can be calculated for
  this choice of mixing function using the formulae given in
  \citet{eliazar09}. The autocovariance function is
  \begin{equation}
    R_{Y}(t) = \frac{\varsigma^2}{2} \left[ E_1(t / \tau_{L}) - E_1(t
      / \tau_{H}) \right],
    \label{eq-Rmixou}
  \end{equation}
  where $\tau_{L} = 1 / \omega_{L}$ and $\tau_{H} = 1 /
  \omega_{H}$ are the maximum and minimum characteristic time scales
  of the mixture, and $E_1(x)$ is an exponential integral of first
  order:
  \begin{equation}
    E_1(x) = \int_{1}^{\infty} \frac{\exp(-xy)}{y}\ dy, \ \ \ x > 0.
    \label{eq-expint}
  \end{equation}
  The variance of the continuous mixed OU process, $Y(t)$, is
  calculated by taking the limit of Equation (\ref{eq-Rmixou}) as $t
  \rightarrow 0$:
  \begin{equation}
    Var[Y(t)] = \lim_{t' \rightarrow 0} R_{Y}(t') =
    \frac{\varsigma^2}{2}\ln\left(\frac{\tau_{L}}{\tau_{H}}\right).
    \label{eq-mixou_var}
  \end{equation}
  Because the individual OU processes involved in the mixture are
  assumed to have zero mean, the mean of $Y(t)$ is $\mu$. 

  The power spectrum for the continuous mixed OU process with
  weighting function given by Equation (\ref{eq-mixfunc}) is 
  \begin{equation}
    P_{Y}(\omega) = \frac{\varsigma^2}{2\pi\omega} \left[
      \tan^{-1}\left(\frac{\omega_{H}}{\omega}\right) -
      \tan^{-1}\left(\frac{\omega_{L}}{\omega}\right)\right].
    \label{eq-Pmixou}
  \end{equation}
  Three regions are of interest,
  \begin{equation}
    P_{Y}(\omega) \sim \left\{ \begin{array}{l}
        \varsigma^2 (\tau_{L} - \tau_{H}) / 2,\ \ \ \omega \ll \omega_{L}\\
        1 / \omega,\ \ \ \omega_{L} \ll \omega \ll \omega_{H} \\
        1 / \omega^2,\ \ \ \omega \gg \omega_{H} \end{array} \right. .
    \label{eq-Pmixou_asympt}
  \end{equation}
  Therefore, the mixed OU process resembles white noise on time
  scales $\delta t \gg \tau_{L}$, `pink' or `flicker' noise on
  time scales $\tau_{H} \ll \delta t \ll \tau_{L}$, and red noise on
  time scale $\delta t \ll \tau_{H}$. For a lightcurve that follows
  a mixed OU process, $\tau_{L}$ represents the maximum time scale
  on which $Y(t)$ is correlated; on time scales longer than
  $\tau_{L}$ the lightcurve `forgets' about its previous
  behavior.

  We can also formulate a mixed OU process with common driving
  noise. This process was studied by \citet{eliazar09}. Physically, we
  might expect this type of process to arise if, for example, the
  fluctuations in the optical lightcurve were the result of reprocessing due to a common source
  which irradiates a range of radii, with the reprocessing time scales
  increasing with increasing radii. Denote this type of mixed OU
  process as $\tilde{Y}(t)$. In this case, the power spectrum using the mixing function defined by Equation (\ref{eq-mixfunc}) is
  \begin{equation}
    P_{\tilde{Y}}(\omega) = \frac{\varsigma^2}{8\pi}
    \left[\ln\left(\frac{\omega^2 + \omega_{H}^2}{\omega^2 +
          \omega_{L}^2}\right)\right]^2 + \frac{\varsigma^2}{2\pi}
    \left[\tan^{-1}\left(\frac{\omega_{H}}{\omega}\right) -
      \tan^{-1}\left(\frac{\omega_{L}}{\omega}\right)\right]^2.
    \label{eq-Pcommon}
  \end{equation}
  The mixed OU process with common driving noise exhibits less
  long-time scale variations than does the mixed OU process with
  independent driving noise, due to the correlation in the processes
  created by a common driving noise.

\clearpage

\begin{deluxetable}{lcccccccccc}
\tablewidth{0pt}
\tabletypesize{\tiny}
%\rotate
\tablecaption{AGN with X-ray Lightcurves Analyzed\label{t-xray_agn}}
\tablehead{
\colhead{Name} 
& \colhead{z}
& \colhead{Lit. $f_{H}$\tablenotemark{a}} 
& \colhead{Lit. PSD Slope\tablenotemark{b}}
& \colhead{Reference} \tablenotemark{c}
& \colhead{$f_{H}$\tablenotemark{d}} 
& \colhead{PSD Slope \tablenotemark{e}}
& \colhead{$\tau_{L}$\tablenotemark{f}}
& \colhead{$\tau_{H}$\tablenotemark{g}}
& \colhead{$\varsigma$\tablenotemark{h}}
& \colhead{RMS Variability} \\
&  & \colhead{Hz} & & & \colhead{Hz} & & \colhead{day} & \colhead{ksec} & 
\colhead{$10^{-2}$ per cent ${\rm sec}^{-1/2}$} & \colhead{per cent rms}
}
\startdata
Fairall 9   & 0.047 & $3.98^{+2.33}_{-2.40} \times 10^{-7}$ & $1.10^{+1.1}_{-0.6}$ & 1 & $2.25^{+154}_{-1.73} \times 10^{-6}$ 
            & $1.58^{+0.25}_{-0.27}$ & $> 211$ & $70.8^{+232}_{-70.0}$
            & $2.72^{+1.44}_{-0.47}$ & $36.8^{+23.1}_{-10.1}$ \\

Akn 564     & 0.025 & $2.4^{+2.3}_{-0.9} \times 10^{-3}$    & $1.2^{+0.2}_{-0.1}$ & 2 & $5.25^{+3.55}_{-2.04} \times 10^{-4}$ 
            & $0.98^{+0.19}_{0.21}$ & $2.27^{+7.39}_{-1.36}$ &
            $0.30^{+0.19}_{-0.12}$ & $92.8^{+9.29}_{-8.24}$ & $27.5^{+1.67}_{-1.52}$ \\

Mrk 766     & 0.013 & $5^{+1}_{-3} \times 10^{-4}$         & 1\tablenotemark{i} & 7 & $2.69^{+1.51}_{-0.94} \times 10^{-4}$
            & $0.96^{+0.10}_{-0.12}$ & $>31.0$ &
            $0.59^{+0.32}_{-0.21}$ & $61.2^{+5.73}_{-5.00}$ & $29.8^{+4.67}_{-3.09}$ \\

NGC 4051    & 0.002 & $2.4^{+13}_{-2.3} \times 10^{-4}$\tablenotemark{j} & $1.1^{+0.1}_{-0.4}$ & 4 & $7.18^{+2.62}_{-1.84} \times 10^{-4}$
            & $1.10^{+0.04}_{-0.04}$ & $> 246$ &
            $0.22^{+0.08}_{-0.06}$ & $120^{+9.53}_{-8.34}$ & $59.0^{+8.17}_{-6.20}$ \\

NGC 3227    & 0.004 & $2.6^{+5.3}_{-2.3} \times 10^{-4}$ & $1.0^{+0.2}_{-0.4}$ & 6 & $3.73^{+2.51}_{-1.44} \times 10^{-5}$
            & $1.20^{+0.08}_{-0.08}$ & $> 287$ &
            $4.26^{+2.67}_{-1.72}$ & $21.8^{+2.68}_{-2.30}$ & $52.2^{+12.2}_{-7.93}$ \\

NGC 5548    & 0.017 & $6.31^{+18.8}_{-5.05} \times 10^{-7}$ & $1.15^{+0.50}_{-0.65}$ & 1 & $7.51^{+8.18}_{-3.33} \times 10^{-7}$
            & $1.22^{+0.29}_{-0.25}$ & $> 221.6$ & $212^{+169}_{-111}$
            & $4.36^{+0.39}_{-0.32}$ & $47.6^{+16.6}_{-8.29}$ \\

NGC 3783    & 0.010 & $6.2^{+40.6}_{-5.6} \times 10^{-6}$  & $0.8^{+0.5}_{-0.8}$ & 5 & $4.00^{+3.24}_{-1.47} \times 10^{-6}$
            & $0.79^{+0.42}_{-0.30}$ & $> 12.6$ &
            $39.8^{+23.3}_{-17.8}$ & $9.02^{+0.90}_{-0.76}$ & $24.4^{+2.39}_{-1.51}$ \\

NGC 5506    & 0.006 & $3.9^{+16}_{-3.8} \times 10^{-5}$ & $1.0^{+0.2}_{-1.0}$ & 6 & $5.25^{+4.27}_{-2.20} \times 10^{-5}$ 
            & $1.11^{+0.16}_{-0.15}$ & $> 18.4$ &
            $3.03^{+2.18}_{-1.36}$ & $18.9^{+2.17}_{-1.90}$ & $24.4^{+4.47}_{-2.20}$ \\

MCG-6-30-15 & 0.008 & $6.0^{+10}_{-5} \times 10^{-5}$ & $0.8^{+0.16}_{-0.4}$ & 3 & $6.80^{+2.12}_{-1.55} \times 10^{-5}$ 
            & $0.70^{+0.10}_{-0.11}$ & $> 36.6$ &
            $2.34^{+0.69}_{-0.56}$ & $42.9^{+3.10}_{-2.87}$ & $27.7^{+1.35}_{-1.12}$ \\

NGC 3516    & 0.009 & $2.00^{+3.01}_{-1.00} \times 10^{-6}$ & $1.10^{+0.4}_{-0.3}$ & 1 & $3.24^{+46.7}_{-2.17} \times 10^{-6}$ 
            & $1.21^{+0.42}_{-0.39}$ & $> 39.07$ &
            $49.1^{+99.1}_{-45.9}$ & $8.28^{+3.07}_{-1.25}$ & $46.4^{+35.7}_{-10.0}$

\enddata

\tablecomments{Error bars denote $90\%$ Confidence Intervals, except
  for values from \citet{mark03} and \citet{mchardy07}, who report
  errors at the $68\%$  level. Lower limits on $\tau_{L}$ are at $99\%$
  confidence.}

\tablenotetext{a}{The location of the high-frequency break previously
  reported in the literature.}  
\tablenotetext{b}{The value of $\alpha, P(\nu) \propto f^{-\alpha}$ in
the intermediate region of the PSD, below the high-frequency break,
previously reported in the literature.}
\tablenotetext{c}{The reference for the literature values of $f_{H}$
  and $\alpha$.}
\tablenotetext{d}{The estimated value of the location of the
  high-frequency break, obtained using our Bayesian implementation of
  the mixture of OU processes model.}
\tablenotetext{e}{Same as the previous column, but for the PSD
  logarithmic slope below the high-frequency break.}
\tablenotetext{f}{The longer characteristic time scale of the
  lightcurve, corresponding to the low-frequency break.}
\tablenotetext{g}{The shorter characteristic time scale of the
  lightcurve, corresponding to the high-frequency break.}
\tablenotetext{h}{The fractional amplitude of the driving noise field for the
  mixed OU process. Since we have normalized the weights to have unit
  norm, $\varsigma^2$ is also proportional to the lightcurve fractional variance on
  time scales short compared to $\tau_{H}$.}
\tablenotetext{i}{\citet{vau03} fixed the intermediate PSD logarithmic
  slope to $\alpha = 1$ for Mrk 766.}
\tablenotetext{j}{The values reported by \citet{mchardy04} are for a
  sharply broken power-law, as the authors did not state confidence
  regions for the bending power-law PSD model.}

\tablerefs{
(1) Markowitz et al. 2003
(2) McHardy et al. 2007
(3) McHardy et al. 2005
(4) McHardy et al. 2004
(5) Summons et al. 2007
(6) Uttley \& McHardy 2005
(7) Vaughan \& Fabian 2003
}
\end{deluxetable}

\begin{deluxetable}{lccccc}
\tablewidth{0pt}
\tablecaption{\emph{MACHO} Sources with Optical Lightcurves that
  Deviate from a Lorentzian PSD\label{t-optical}}
\tablehead{
\colhead{\emph{MACHO} ID}
& \colhead{RA (J2000)}
& \colhead{Dec (J2000)} 
& \colhead{z}
& \colhead{V-band PSD Slope \tablenotemark{a}}
& \colhead{R-band PSD Slope}
}
\startdata

2.5873.82       & 05 16 28.78 & -68 37 02.38 & 0.46 & $1.42^{+0.05}_{-0.06}$ & $1.35^{+0.05}_{-0.06}$ \\
9.5484.258     & 05 14 12.05 & -70 20 25.64 & 2.32 & $1.42^{+0.10}_{-0.10}$ & $1.63^{+0.29}_{-1.40}$ \\
13.5717.178   & 05 15 36.02 & -70 54 01.65 & 1.66 & $1.40^{+0.09}_{-0.10}$ & $1.50^{+0.23}_{-0.20}$ \\
68.10968.235 & 05 47 45.13 & -67 45 5.745 & 0.39 & $1.15^{+0.17}_{-0.31}$ & $0.98^{+0.17}_{-0.30}$ \\
69.12549.21   & 05 57 22.41 & -67 13 22.16 & 0.14 & $1.57^{+0.11}_{-0.12}$ & $1.82^{+0.14}_{-0.21}$ \\
70.11469.82   & 05 50 33.31 & -66 36 52.96 & 0.08 & $1.69^{+0.24}_{-1.30}$ & $1.60^{+0.21}_{-0.22}$ \\
82.8403.551   & 05 31 59.66 & -69 19 51.12 & 0.15 & $1.10^{+0.06}_{-0.11}$ & $1.47^{+0.31}_{-0.86}$ \\

\enddata

\tablecomments{Error bars denote $90\%$ Confidence Intervals.}

\tablenotetext{a}{The value of $\alpha, P(f) \propto 1 / f^{\alpha}$.}

\end{deluxetable}

\end{document}